%% file: Collaposer.tex
\PassOptionsToPackage{table}{xcolor}
\documentclass[sigconf,screen,authorversion]{acmart}
\AtBeginDocument{%
  }

\setcopyright{acmlicensed}
\copyrightyear{2026}
\acmYear{2026}
\setcopyright{cc}
\setcctype{by}
\acmConference[CHI '26]{Proceedings of the 2026 CHI Conference on Human Factors in Computing Systems}{April 13--17, 2026}{Barcelona, Spain}
\acmBooktitle{Proceedings of the 2026 CHI Conference on Human Factors in Computing Systems (CHI '26), April 13--17, 2026, Barcelona, Spain}
\acmPrice{}
\acmDOI{10.1145/3772318.3791160}
\acmISBN{979-8-4007-2278-3/2026/04}
\acmSubmissionID{3116}

\input{meta/commands}

\begin{document}

\title{\system{}: Transforming Photo Collections into Visual Assets for Storytelling with Collages}

\input{meta/authors}

\begin{abstract}
    \input{meta/abstract}
\end{abstract}

\input{meta/keywords}


\begin{teaserfigure}
  \includegraphics[width=\textwidth]{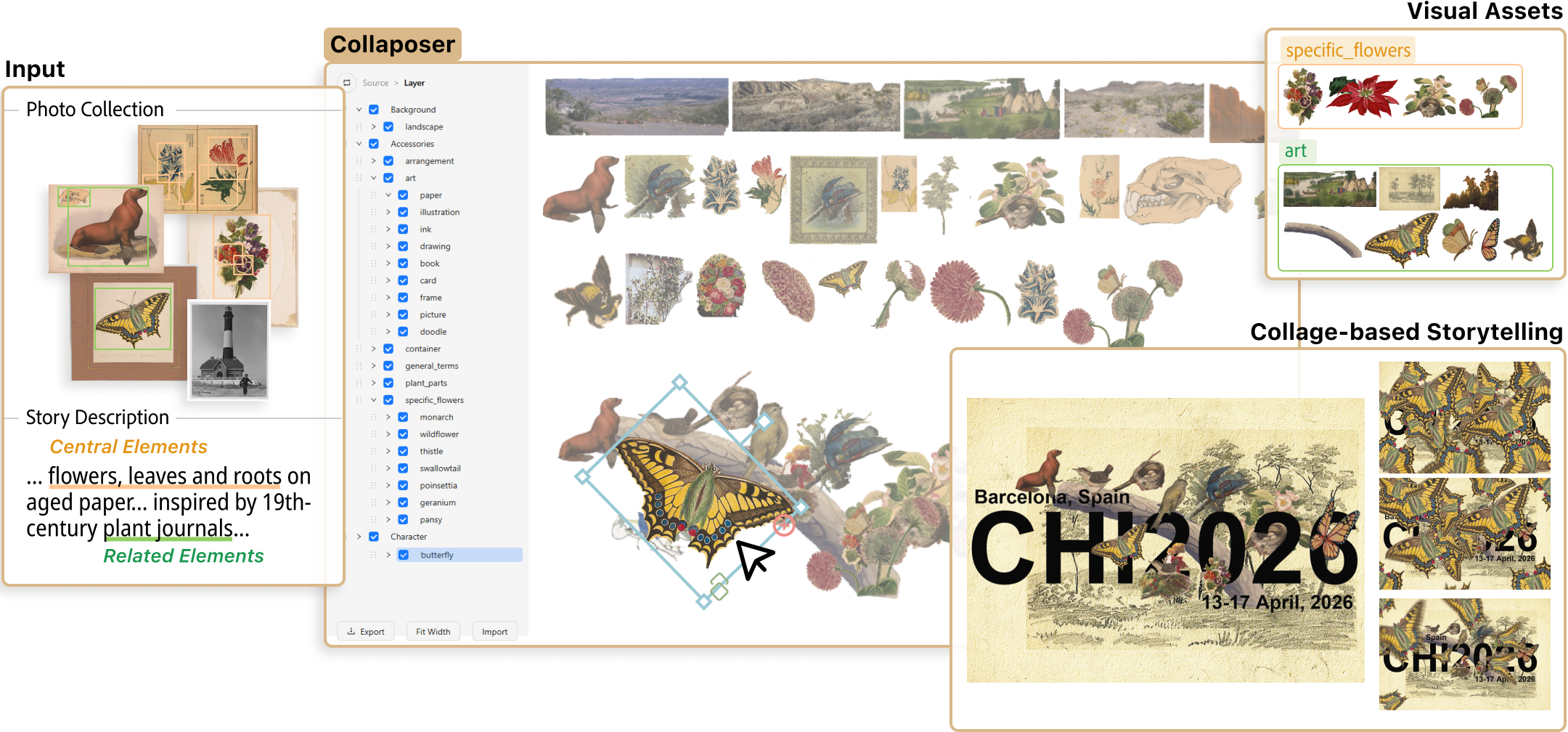}
  \caption{Based on the input photo collection and story descriptions, \system supports visual assets preparation for collage-based storytelling by automatically selecting, cutting out, and presenting diverse visual assets.
  Given a curated overview of the extracted visual elements, users can compose them into a static collage, and animate them to create expressive visual stories.
  Photo collections included are illustrative examples from public-domain image collections~\cite{congress, nasa}.
  }
  \Description{collaposer system overview}
  \label{fig:teaser}
\end{teaserfigure}


\maketitle

\input{sections/01_introduction}
\input{sections/02_relatedWork}
\input{sections/03_formativeStudy}
\input{sections/04_collaposer}
\input{sections/05_usageScenario}
\input{sections/06_evaluation}
\input{sections/07_Results}
\input{sections/08_discussion}
\input{sections/09_limitationsFutureWork}
\input{sections/10_conclusion}
\balance

\begin{acks}
This work was partially supported by the Hong Kong RGC GRF grant TRS T22-607/24N.
We extend our sincere gratitude to animator Lucas Mariano, whose inspiring collage animations and valuable feedback contributed to the iterative development of the system.
We also wish to thank Zhicheng Li, Wenshuo Zhang, Runhua Zhang, Shuchang Xu, Leixian Shen, and Prof. Jeffrey V. Nickerson for their generous support and discussions related to this work.
We sincerely appreciate all participants for their time and effort.
Lastly, we are grateful to the reviewers for their constructive suggestions.
\end{acks}

\bibliographystyle{ACM-Reference-Format}
\bibliography{reference}

\appendix
\input{sections/11_appendix}

\end{document}

%% file: meta/commands.tex
\usepackage{booktabs}
\usepackage{enumitem}
\usepackage{xspace}
\usepackage{listings}
\usepackage{tabularx}
\usepackage{listings}
\setlist[itemize]{left=0pt}

\newcommand{\RR}[1]{\textcolor{black}{#1}}
\newcommand{\system}{Collaposer\xspace}
\newcommand{\baselineA}{\textit{Ablated-Select}\xspace}
\newcommand{\baselineB}{\textit{Ablated-Present}\xspace} 
\newcommand{\user}{Lucas\xspace}
\definecolor{purpletext}{HTML}{624aa1}
\definecolor{bluebright}{HTML}{4a9aef}
\definecolor{lightgray}{HTML}{bfbfbf}

\newcommand{\shelly}[1]{\textcolor{teal}{[Shelly] #1}}
\newcommand{\anyi}[1]{\textcolor{red}{[Anyi] #1}}
\newcommand{\jiaju}[1]{\textcolor{blue}{[Jiaju] #1}}
\newcommand{\zoe}[1]{\textcolor{orange}{[Zoe] #1}}

\newcommand{\p}[1]{%
  \par\addvspace{0.3\baselineskip}%
  \par\noindent\textbf{#1}
}

\AtBeginDocument{

}

\lstdefinestyle{liststyle}{       
backgroundcolor=\color[rgb]{0.95,0.95,0.95},
  numberstyle=\tiny\color[rgb]{0.5,0.5,0.5},
  basicstyle=\ttfamily\footnotesize,
  breakatwhitespace=false,         
  breaklines=true,                 
  captionpos=b,                    
  keepspaces=true,                 
  numbers=left,                    
  numbersep=5pt,                  
  showspaces=false,                
  showstringspaces=false,
  showtabs=false,                  
  tabsize=2,
  frame=ltb,
  framerule=0pt,
}

%% file: meta/authors.tex


\author{Jiayi Zhou}
\email{jzhoudp@connect.ust.hk}
\orcid{0000-0003-4669-4872}
\affiliation{
  \institution{The Hong Kong University of Science and Technology}
 \city{Hong Kong SAR}
 \country{China}
}

\author{Liwenhan Xie}
\email{lxieai@connect.ust.hk}
\orcid{0000-0002-2601-6313}
\affiliation{
  \institution{The Hong Kong University of Science and Technology}
  \city{Hong Kong SAR}
  \country{China}
}

\author{Jiaju Ma}
\email{jiajuma@stanford.edu}
\orcid{0000-0003-2880-8506}
\affiliation{
  \institution{Stanford University}
  \city{Stanford, CA}
  \country{USA}
}

\author{Zheng Wei}
\email{zwei302@connect.ust.hk}
\orcid{0000-0001-7444-2547}
\affiliation{
  \institution{The Hong Kong University of Science and Technology}
  \city{Hong Kong SAR}
  \country{China}
}

\author{Huamin Qu}
\email{huamin@cs.ust.hk}
\orcid{0000-0002-3344-9694}
\affiliation{
  \institution{The Hong Kong University of Science and Technology}
  \city{Hong Kong SAR}
  \country{China}
}

\author{Anyi Rao}
\email{anyirao@ust.hk}
\orcid{0000-0003-1004-7753}
\affiliation{
  \institution{The Hong Kong University of Science and Technology}
  \city{Hong Kong SAR}
  \country{China}
}

\renewcommand{\shortauthors}{Zhou, et al.}

%% file: meta/abstract.tex
Digital collage is an artistic practice that combines image cutouts to tell stories. However, preparing cutouts from a set of photos remains a tedious and time-consuming task.
A formative study identified three main challenges: 1) inefficient search for relevant photos, 2) manual image cutout, and 3) difficulty in organizing large sets of cutouts.
To meet these challenges and facilitate asset preparation for collage, we propose \system, a tool that transforms a collection of photos into organized, ready-to-use visual cutouts based on user-provided story descriptions.
Collaposer tags, detects, and segments photos, and then uses an LLM to select central and related labels based on the user-provided story description.
\system presents the resulting visuals in varying sizes, clustered according to semantic hierarchy.
Our evaluation shows that Collaposer effectively automates the preparation process to produce diverse sets of visual cutouts adhering to the storyline, allowing users to focus on collaging these assets for storytelling.

%% file: meta/keywords.tex
\ccsdesc[500]{Human-centered computing~Interactive systems and tools}
\ccsdesc[500]{Applied computing~Media arts}

\keywords{Collage, Storytelling, Human-AI Collaboration, Creativity Support Tools, Visual Assets Management}

%% file: sections/01_introduction.tex

\section{Introduction}

Digital collage pieces together fragments of photographs, illustrations, and video stills into a story that feels layered and alive.
It has been an enduring expressive art form prevalent in social media platforms~\cite{mcleod2011cutting}, \RR{such as the animated collage feature in Instagram Story~\cite{haberman2025instagramCollage}}, online ad~\cite{PrandBonJourney}, and digital zines~\cite{lyons2021embracing}.
However, this powerful storytelling method is frequently hindered by the significant preparatory labor involved in converting potential sources into visual assets.
In practice, creators struggle to translate their story intent into effective search queries.
They must sift through vast and disorganized collections of images and isolate visual elements down to individual body parts\footnote{Throughout this paper, we use \textit{element} to refer to a single object extracted from an image, and \textit{assets} to denote curated visual materials prepared for further editing and composition.} before composition can begin.
As a result, the creators often spend a disproportionate amount of time on the mechanics of asset preparation, rather than iterating on the stories they are telling.

\RR{Many systems help users discover or organize visual materials from large collections for photo summarization ~\cite{c2-collage-authoring, c3-narrative-collage, c5-autocollage, c10-pan2019content}, design reference search~\cite{wang2025aideation, rao2024scriptviz, Koch2020SemanticCollage}, and visual assets management~\cite{V6-10.1145/1520340.1520602, V7-10.1145/1520340.1520528, V5-10.1145/3606030}.
Recent advances in large–language models further facilitate flexible natural-language-mediated editing of visual assets~\cite{T14-lave-video-editing}.
While the systems support idea remixing and style exploration, they largely operate at the level of whole images or clips, while Collage-based storytelling requires fine-grained visual elements.
Therefore, visual asset preparation in collage creation remains a tedious workflow that often requires labor-intensive processes such as searching for candidate images, extracting clean cutouts, organizing them, and assigning intuitive names.
To better understand the challenges encountered by practitioners, we conducted a formative study with experienced collage creators (\(N{=}12\)), which uncovered three primary challenges: 1) gaps between loosely specified story intent and asset retrieval, 2) the labor of segmentation and part-level parsing that interrupts momentum, and 3) the burden of asset organization needed to stay oriented.}

To address the challenges faced by collage creators, we introduce \system, an interactive system for collage asset curation.
Our goal is to uncover story-element associations beyond literal mentions, provide semantic grouping for navigation, and enable automated extraction of meaningful elements.
Given a collection of photos and a brief story description, \system automatically curates and presents a layered set of ready-to-use visual elements.
The pipeline first labels, detects, and performs instance segmentation to extract candidate elements into a visual element library.
Through a large language model (LLM), it then selects labeled elements, both explicitly mentioned (central elements) and implied from the context (related elements); classifies them into a pre-defined ontology of characters, backgrounds, and accessories; clusters them into a semantic hierarchy; and ranks them with a selection score based on story consistency, diversity, and resolution. 
For characters, the system further parses parts and keypoints to enable rich post-editing, such as deformation and articulated animation.
All assets can be exported as structured, layered files.
An interface with a linked tree view and canvas presents an immediate overview of options without renaming or re-cutting.





To validate our approach, we conducted a user study  (\(N{=}12\)) evaluating \system's selection quality, user experience, and impact on the creative process.
Its performance was benchmarked against two ablated baselines, which respectively removed its core selection reasoning and presentation structure.
Participants of varying expertise each created three static collage stories.
\system outperformed both baselines in terms of asset–story consistency, selection diversity, presentation, and usability; participants also required fewer prompt attempts on average.
Under our 15-minute task budget, all participants achieved a satisfactory result in a single pass with Collaposer.
Qualitative feedback indicated that integrating selection, organization, and segmentation allowed participants to spend time composing scenes rather than hunting, naming, and cutting materials.

In summary, we make the following contributions: (1) a formative study that identifies and summarizes the practices and challenges of asset preparation in collage-based storytelling; (2) \system, a pipeline coupling open-set detection and instance segmentation with LLM-based association, semantic grouping, and character part parsing to produce ready-to-use assets; and (3) a user study evaluating the usefulness and effectiveness of \system with users of varying expertise levels.

%% file: sections/02_relatedWork.tex
\section{Related Work}
\RR{We review related literature that addresses visual assets preparation for collage-based storytelling and approaches that attend to design material extraction and visual assets management in general visual content creation tools.}



\subsection{\RR{Visual Assets Preparation for Collage-based Storytelling}}
\RR{Collage is a visual storytelling practice that recontextualizes fragments of existing visual materials~\cite{Barron2023Collage, wagner2015essence}.
Unlike the posters or video shorts, which often incorporate existing fragments as decorative assets or background footage to support a predetermined message, collage foregrounds the act of recombination itself: visual fragments are not secondary supports but the primary building blocks of meaning.
By rearranging the stated fragments, collage-based storytelling can detach materials from their original context to create new narratives, offering an accessible yet expressive visual storytelling strategy that does not require specialized drawing or filmmaking skills~\cite{Diggs2015TherapeuticCollage, Leland2011Collage}.
Commercial tools that collage creators frequently use (e.g.,  Procreate~\cite{Procreate}, Adobe Photoshop~\cite{AdobePhotoshop}) enable flexible composition, precise masking, and a rich visual manipulation.
However, the commercial tools either assume that creators have visual assets available for use or offer only a small set of built-in elements with limited customization.
Preparing the visual assets remains a demanding process~\cite{Gowrley2024}.}

\RR{
Existing academic tools partially support asset preparation for collage-based storytelling, often limiting either the story scope or the available materials.
Research on photo summarization~\cite{c3-narrative-collage, c5-autocollage, c9-wu2016very, c10-pan2019content} prioritizes the automatic extraction and composition of visually salient content from personal albums, optimizing for criteria such as spatial constraints, visual aesthetics, representativeness, and narrative coherence.
For example, Pan et al.~\cite{c10-pan2019content} propose a visual summarization algorithm that selects and arranges images based on content diversity, conciseness, and aesthetic considerations.
However, the systems cannot support flexible creation of new stories that extend beyond the original photo content, as visual elements remain bundled within each photograph.
A second line of work~\cite{c6-physical-collage, c7-jo2024collagevis, c8-panjwani2010collage} sidesteps preparation issues by collecting assets through taking instant photographs and recording videos.
While ensuring consistency between story and material, it places substantial burden on creators and does not assist with reusing existing image collections.
A recent workshop on AI-supported digital collage creation highlights the promise of generative assistance for tasks such as theme ideation, asset generation, image editing, and interpretation~\cite{Luo2025AIcollage}, but still relies on conventional search engines for image retrieval.
To enable flexible storytelling that moves beyond while being grounded in existing materials, we propose a story-driven asset preparation pipeline that automatically extracts, selects, and organizes object-level visual assets from photo collections.}

\subsection{Design Material Extraction}
\RR{Creators collect relevant design materials for two intertwined purposes:
to gather visual references that support idea remixing, stylistic exploration, and contextual grounding~\cite{Herring2009getting, Keller2006collections, sharmin2009understanding};
and to obtain materials that can be directly incorporated into the composition~\cite{T14-lave-video-editing, benharrak2025historypalette}.
During the extraction process, creators seek accurate matches that align with their current vision, yet also value serendipitous discoveries that spark unexpected creative possibilities.}

\RR{A growing body of tools explores cross-modality retrieval from large visual corpora to help creators efficiently obtain relevant design materials~\cite{cai2023designaid, kang2021metamap, Koch2020SemanticCollage, choi2025expandora, Ritchie2011dtour, Bunian2021VINS, T14-lave-video-editing}.
For example, AIdeation~\cite{wang2025aideation} processes user-provided images and instructions via an Idea Generation LLM to produce keywords, which users can select to retrieve relevant images through search engines.
ScriptViz~\cite{rao2024scriptviz} preprocesses a large movie database with scene-, shot-, and frame-level attributes, and constructs queries that combine user-specified fixed and variable attributes.
SemanticCollage~\cite{Koch2020SemanticCollage} enables multimodal search using both images and keywords.
Most closely related to our work, LAVE~\cite{T14-lave-video-editing} supports natural language-based video retrieval by embedding both video narrations and user queries into a shared semantic space.
Similarly, WYTIWYR~\cite{xiao2023wytiwyr} supports intent-aware chart retrieval by extracting key visual attributes from a query chart image (e.g. layout, chart type, and color scheme) and incorporating user-specified textual intent into a multimodal embedding.
The systems generally return design materials at the level of whole images or clips.
However, in collage-based storytelling, design materials are discrete visual elements extracted from existing images that can be recombined into new stories.
Consequently, collage creators often spend substantial effort manually extracting usable components or crafting new materials to suit their storytelling goals.
Our system enables object-level asset extraction by performing instance segmentation to preprocess images into reusable visual elements and automatically generate semantic tags without requiring predefined attribute vocabularies.
Furthermore, by incorporating user intent via natural-language queries and supplementing user expression with LLM-based semantic association, \system expands retrieval coverage and encourage serendipitous discoveries.
Together, the above-mentioned capabilities complement existing tools in extraction fidelity and streamlines the conversion of story ideas into sufficient visual assets.}

\subsection{Visual Assets Management for Reuse}
\RR{
Visual assets management, including organizing, cataloging, and retrieving large media collections is labor-intensive~\cite{Austerberry2012assetsManagement}.
As digital visual content continues to grow explosively, this challenge becomes increasingly salient while creators gain unprecedented opportunities to reuse digital materials~\cite{Liu2021Reuse, xie2025datawink}.
Commercial tools~\cite{wiki:AdobeBridge, CanvaTemplates, figmaLibraries} embed functions of manual categorizing, tagging, sharing, and version control of design assets.
Figma Libraries~\cite{figmaLibraries}, for instance, enable collaborators to publish shared design assets, including UI components, styles, and variables, that can be reused across projects.
To support users discover, organize, and repurpose visual materials, academic systems have explored automatic visualization~\cite{V1-demian2009effective, V3-10.1145/1978942.1979171} and recommendation~\cite{V4-10.1145/2072298.2072451, V2-10.1145/3336191.3371866}.
Jones et al.'s tool~\cite{V5-10.1145/3606030} automatically selects and visualizes exemplars to facilitate understanding and sharing of large web archives.
MetaMap~\cite{kang2021metamap} introduces a mind-map visualization of examples to recommend and assist visual metaphor exploration.
StoryImaging~\cite{V4-10.1145/2072298.2072451} clusters and ranks images harvested from the Web to enhance textual stories, while Chowdhury et al.~\cite{V2-10.1145/3336191.3371866} embed images from personal albums into relevant textual contexts.
Tagging is an effective method to establish connections across visual assets~\cite{V6-10.1145/1520340.1520602, V7-10.1145/1520340.1520528}.
Souvenirs~\cite{V7-10.1145/1520340.1520528} allows users to create a tangible link between digital photo sets and physical memorabilia for in-home photo sharing.
StoryTags~\cite{V6-10.1145/1520340.1520602} supports abundant, high-quality tagging by encouraging users to tell a story about each photo.
More closely related to collage, HistoryPalette~\cite{benharrak2025historypalette} supports exploration of object-level past generated design alternatives organized by position, concept, and time, allowing users to preview them in their original position.
Building on these visual assets management approaches, \system extracts and organizes object-level visual assets into a semantic hierarchy aligned with the user-provided story description.
By grounding asset preparation in narrative intent, it structures assets as manipulable components that creators can select and recontextualize to support story construction.
The top-down semantic organization of scene components (e.g., characters, backgrounds, accessories) creates a rich creation space for storytelling.
As a result, our pipeline shifts visual assets management from a retrieval-oriented task to a constructive, story-driven process.}

%% file: sections/03_formativeStudy.tex
\begin{table*}[t]
\setlength{\tabcolsep}{6pt}
\centering
\caption{Background of the collage creators in the formative study: their occupation, base region, years of experience in creating collage stories, and collaging frequency.}
\label{t:participants}
\begin{tabular}{lllrl}
\toprule
\textbf{ID} & \textbf{Occupation} & \textbf{Base Region} & \textbf{Experience} & \textbf{Frequency} \\ 
\midrule
$E_{1}$ & 2D animation artist & Hong Kong SAR & 18 & weekly \\
$E_{2}$ & 2D animator & China & 6 & daily \\
$E_{3}$ & graphic designer & United States & 6 & daily \\
$E_{4}$ & motion designer & Brazil & 14 & daily \\
$E_{5}$ & 2D animator & Kazakhstan & 9 & daily \\
$E_{6}$ & video journalist & Germany & 15 & monthly \\
\bottomrule 
\end{tabular}
\end{table*}

\section{Formative Study}
To inform the design of \system, we conducted a formative study with collage creators. Our goal was to understand their existing workflows and pain points, focusing on (1) how they currently prepare assets for storytelling and (2) what challenges present opportunities for automation and AI support.

\subsection{\RR{Methods}}
\RR{We adopted a qualitative, interview-based approach combining contextual inquiry and artifact walkthrough~\cite{Shi2025Brickify}. 
This method allowed us to observe creators’ actual workflows and decision-making processes while grounding the discussion in concrete project artifacts.
We conducted semi-structured interviews with professional collage creators and analyzed the collected data using thematic analysis~\cite{kiger2020thematic} to identify recurring patterns and challenges.}

\subsection{Participants and Procedure}
We reached out via email to creators whose collage work is featured on social media and recruited six participants (denoted as $E_1$–$E_6$) \RR{with 6–18 years of experience in collage-based storytelling. Their work spans award-winning animation ($E_1$, $E_4$), game concept art ($E_2$), high-visibility online content (e.g., a YouTube short with 410K views by $E_3$), advocacy short film ($E_5$), and socially engaged short films ($E_6$).
Together, they bring substantial domain expertise and diverse professional backgrounds, as summarized in \autoref{t:participants}.}

Before the interview, we required participants to prepare intermediate artifacts of a previous collage piece for revisiting later.
This artifact-based approach allowed us to ground the discussion in concrete evidence of their process, minimizing recall bias.
The artifacts included scripts, mood boards, static and animated storyboards, reference videos, featured frames, Photoshop projects, After Effects projects, and asset folders.
The interviews began with background questions about the collage-based storytelling piece and the participant's role in creating it. 
Participants were then asked to revisit the creation workflow by demonstrating the intermediate artifacts and manipulating them on the software they used for creation.
Focusing on the preparation of visual assets, we carefully examined \textbf{when} the creators began collecting the assets, \textbf{what} are the frequent sources they turn to, \textbf{what} types of assets they were seeking, \textbf{how} they presented them, and \textbf{how} the preparation supported editing in the subsequent tasks.
The creators reflected on both practices that supported the preparation and the challenges remained unresolved.

The interviews were conducted through Zoom, with an average duration of 69 minutes (SD=14, Range=56--100).
Each participant was compensated US\$50 per hour via gift cards.
The price was based on the average salary rate for professional animators.
During the interviews, both audio and screen sharing were recorded.
To analyze the data collected from the semi-structured interviews, we reviewed the recordings and took detailed notes, organizing the \RR{identified challenges} and the corresponding practices of the creators.
\RR{We then conducted a thematic analysis~\cite{kiger2020thematic} to iteratively group related observations into higher-level themes, refining categories through multiple rounds of discussion among the authors.}\footnote{The study protocol was approved by the institutional review board of the authors’ institution.}

\begin{figure*}[t]
    \centering
    \includegraphics[width=1\linewidth]{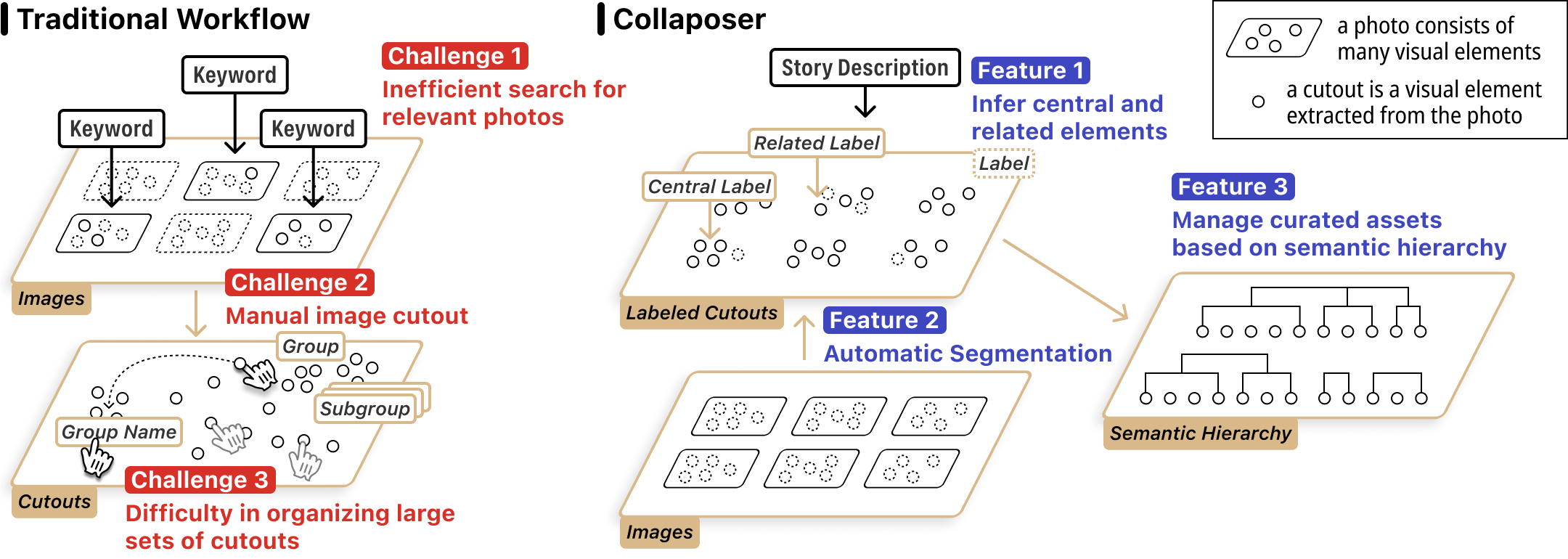}
    \caption{\RR{A comparison between manual workflows and our approach for asset preparation in collage-based storytelling.
    Our formative study identified three challenges in manual workflows: inefficient keyword-based photo search, tedious manual image cutout, and difficulty in organizing large collections of cutouts through manual grouping and naming. \system introduces three corresponding AI-driven features that alleviate the challenges: inference of central and related elements based on the story, automatic instance segmentation, and asset management with semantic hierarchy.}}
    \Description{Challenges in traditional asset preparation workflows and how Collaposer’s design features leverage AI to address them.}
    \label{fig:comparison}
\end{figure*}
\subsection{Identified Challenges}
Our analysis of the study data identified the following recurring challenges (see \autoref{fig:comparison}) that participants reported as impeding efficiency in visual assets preparation for collage-based storytelling.

\p{C1: Inefficiency in searching for relevant
photos.}
Participants struggle to articulate their selection intentions in ways that search systems can support.
They had to translate representations of their story ideas—such as low-fidelity storyboards ($E_{1-2}$), scripts ($E_{6}$), textual descriptions ($E_{2}$, $E_{4}$), and themes ($E_{3}$, $E_{6}$)—into fragmented keyword queries to retrieve visual elements.
However, such queries rarely capture all the necessary elements for the scene, so they had to alternate between material collection and manipulation to compensate elements they had overlooked.
For example, $E_{6}$ spent approximately a month exploring the Library of Congress~\cite{congress} prior to visual story construction, seeking to identify more elements that could scaffold her creative process.
Describing envisioned visuals with precise queries posed additional challenges.
As $E_{2}$ explained, ``{\it I often spend a long time searching for an image that fits the scene I envision by testing many different keyword combinations.}''
Since creative intent often emerges through reflective conversation with design materials rather than being fully predetermined~\cite{Schon1992, creativeDesign},
participants deliberately expanded their search with diverse alternatives and welcomed serendipitous discoveries.
$E_{4}$ collected multiple images of a TV host with different expressions and gestures, while $E_{2}$ explored a range of natural elements to enrich spring scenes.
$E_{6}$ remarked, ``{\it There is always something interesting, maybe a little weird that I haven't thought about.}''
This pursuit of diversity and serendipity conflicts with the precision of keyword-based search, necessitating the planning of multiple queries and active browsing.

\p{C2: Tedious cutout of visual elements.}
Extracting visual elements from embedded backgrounds is often necessary and involves segmentation at multiple granularities.
To support visual effects, movements, and deformations, complete character elements are sometimes insufficient ($E_{1-5}$).
Participants separate body parts to enable keyframe-based editing of each individual segment (rotation, transformation, scaling, placement), a process that is both repetitive and laborious.
As $E_{3}$ complained, ``{\it It's always the same. The foot, lower leg, and upper leg form a leg, and I have to repeat the composition every time.}''
Another strategy is to prepare specialized charts for certain body parts that change over time (e.g., mouth charts for facial expressions, $E_{4}$), adjusting control points to animate movements.
During manipulation, participants frequently realize that new images are needed to support the story ($E_{1-6}$) or that the existing segmentation is insufficient for the intended movements ($E_{1-5}$).
The repeated alternation between cutting, searching, and manipulating assets disrupts creative momentum~\cite{Kreminski2024Intent}.

\p{C3: Difficulty in managing visual elements.}
Whether they work independently ($E_{1-3}$) or collaboratively ($E_{4-6}$), participants emphasized the importance of maintaining organized visual assets.
As $E_{5}$ stressed, ``{\it You have to make everything very, very organized.}''
Managing large collections requires a semantic hierarchy to support efficient retrieval, often through strict adherence to layered clustering and clear naming.
Folders in local storage, compositions in Adobe After Effects, and groups in software such as Photoshop, Illustrator, and Procreate all reflect this practice.
As $E_{3}$ noted, ``{\it Categorizing and naming everything takes a lot of time and gets repetitive,}'' yet participants agreed that clear, unique, and descriptive names are critical for navigation.
As $E_{4}$ explained, ``{\it I see the cluster name, and I know which group of objects it refers to.}''

\subsection{Design Goals}
Based on the challenges identified in our formative study, we derived the following design goals to inform the development of \system (\autoref{fig:comparison}).
To achieve the goals, we seek to harness AI to offload tedious and inefficient tasks for human creators, allowing them to focus on higher-level creative decisions.

\p{DG1: Enhance selection efficiency by associating related visual elements.}
Participants often struggle to fully articulate their selection intentions, relying on fragmented keyword queries that rarely capture all necessary elements (C1).
Precise keyword-based search is incompatible with efforts to seek diverse alternatives and embrace serendipitous stimuli (C1).
This reveals a gap between creators’ selection intentions and systems’ retrieval capability.
To bridge the gap, systems should supplement limited selection expressions by uncovering associations to related visual elements.
Such an association mechanism would allow creators to assemble sufficient visual elements through loosely specified descriptions.

\p{DG2: Enable automated extraction through segmenting meaningful visual elements.}
Pre-processing requires extensive manual segmentation and rigging, forcing creators to repeatedly alternate between cutting, searching, and manipulating assets—a cycle that disrupts creative momentum (C3).
To address this, systems should automate extraction of visual elements within search workflows, minimizing repetitive manual effort.
By integrating segmentation into the search process, the goal is to let creators focus on storytelling and exploration rather than mechanical asset preparation.

\p{DG3: Facilitate asset management via semantic reasoning over curated cutouts.}
Existing systems largely depend on manual categorization and naming, a process that is repetitive and time-consuming, yet essential for efficient navigation (C2).
As such, systems should provide structured clustering mechanisms that organize curated visual elements into meaningful groups (e.g., characters, backgrounds, and accessories).
The goal is to establish a semantic hierarchy that enables creators to efficiently locate, manage, and connect visual assets.





%% file: sections/04_collaposer.tex
\section{\RR{\system: System Design and Implementation}}
In this section, we describe the rationale and implementation of our story-driven visual assets preparation pipeline (\autoref{sec:pipeline}).
We then describe the design decisions behind \system's user interface, which instantiates this pipeline (\autoref{sec:interface}).

\begin{figure*}[t]
    \centering
    \includegraphics[width=1.0\linewidth]{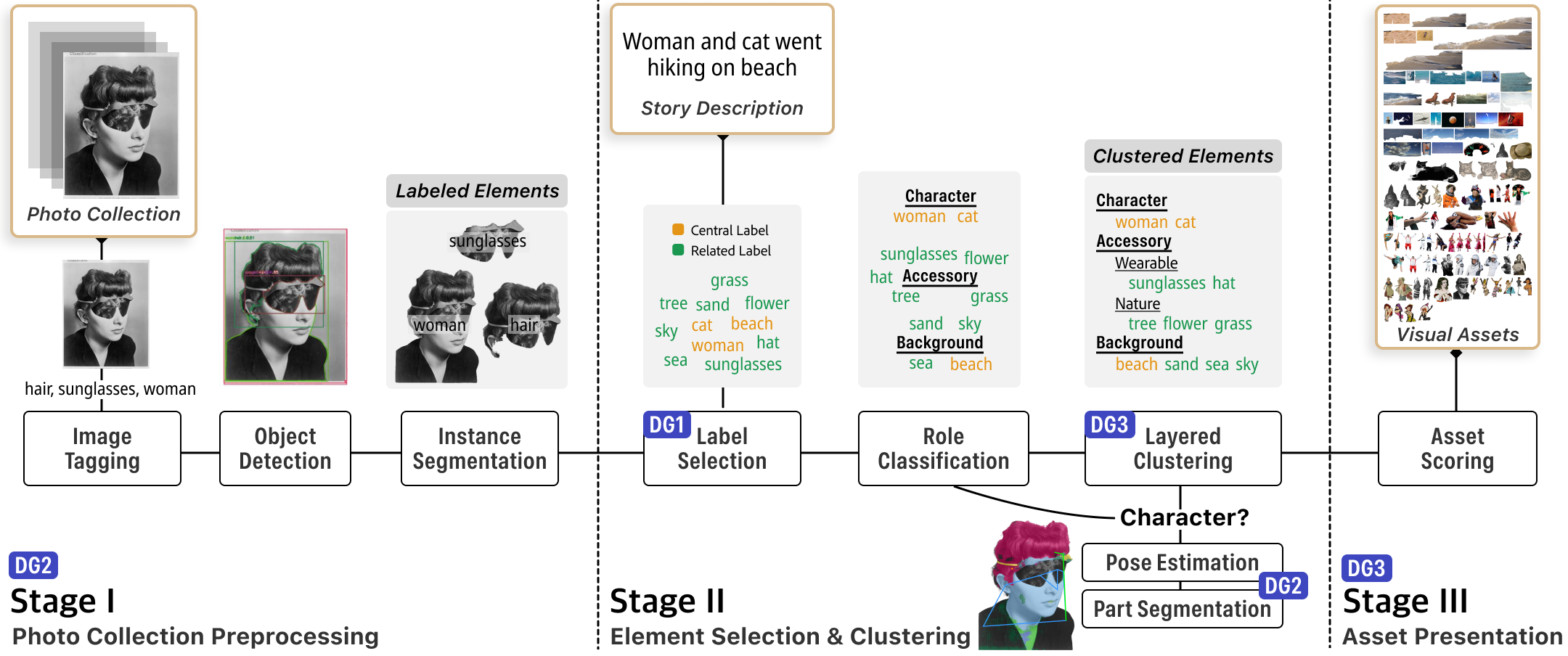}
    \caption{
    Our pipeline consists of three stages. 
    The inputs include an image collection and a story description.
    In Stage I, valid visual elements are trimmed out and tagged with an object name. 
    In Stage II, visual elements relevant to the story are selected and clustered into semantic groups.
    The elements classified as characters undergo part segmentation and pose estimation for later manipulation. 
    In Stage III, the visual assets are presented in a compact view to facilitate navigation and composition.}
    \Description{Our pipeline consists of three stages. 
    The inputs include an image collection and a story description.
    In Stage I, valid visual elements are trimmed out and tagged with an object name. 
    In Stage II, visual elements relevant to the story are selected and clustered into semantic groups.
    The elements classified as characters undergo part segmentation and pose estimation for later manipulation. 
    In Stage III, the visual assets are presented in a compact view to facilitate navigation and composition.}
    \label{fig:pipeline}
\end{figure*}
\subsection{Story-driven visual assets preparation}
\label{sec:pipeline}

\noindent
Our pipeline (see \autoref{fig:pipeline}) takes as input a set of images and a textual description of the desired story content.
It is split into three stages:  photo collection preprocessing (Stage I), elements selection and clustering (Stage II), and asset presentation (Stage III).

\subsubsection{Stage I: Photo Collection Preprocessing}
Collaposer automates the cutout extraction process (\textbf{DG2}) by sequentially processing each photo in the collection, applying state-of-the-art computer vision models to label, detect, and segment visual elements.
Stage I produces a set of cutouts extracted from the photos, along with labels aggregated across images.
Each cutout is stored as an RGBA image named after its corresponding label and assigned a unique ID to distinguish multiple elements sharing the same label.
Since a cutout can be associated with multiple labels, it may appear multiple times in the asset library under different labels.
The resulting semantic labels then serve as the basis for selection and clustering in Stage II. Stage I consists of the following steps:

\p{Image Tagging.}
The first step aims to extract semantic tags across images, serving as representative concepts for visual elements.
We use RAM~\cite{zhang2023recognize} to tag images in the photo collection.
We choose RAM because its strong zero-shot tagging ability covers diverse objects and scenes beyond fixed category sets, making it well-suited for open-ended collage asset preparation.
Semantic labels encompassing objects, scenes, attributes, and actions are generated for each image.
As attributes and actions alone can not represent visual elements, we then filter the attribute and action tags using Parts-of-Speech Tagging, retaining only the objects and scenes labels to represent visual elements.

\p{Object Detection.}
The next step is to localize visual elements that match the tags within each photo.
The resulting grounding boxes serve as guidance for image segmentation, ensuring that cutouts accurately correspond to their semantic labels.
We utilize Grounding DINO's open-set object detection ability~\cite{liu2023grounding} to detect bounding boxes in the photo for each tag.
Labels and visual elements do not correspond one-to-one.
We locate all elements in an image that match each label, so a single label can correspond to multiple visual elements, and a single element can match multiple labels.
For example, there might be multiple flowers in one photo, or a single visual element could be associated with both the ``spaceship'' and ``space'' labels.
The issue of duplicate detection of visual elements is addressed in Stage III.

\p{Instance Segmentation.}
In the third step, cutouts are generated for each visual element and associated with their respective labels.
With bounding boxes as input prompts for instance segmentation, we subsequently extract visual elements from an image utilizing SAM's zero-shot promptable segmentation capability~\cite{Kirillov_2023_ICCV_SAM}.
Visual elements are cut out by setting the RGBA channels to 0 according to the masks provided by RAM and cropping based on the bounding rectangles defined by the mask edges.

In our original design, we attempted to filter out photos that appeared to conflict with user input; however, it risked discarding images containing elements that could enhance the narrative.
For example, while a woman in a swimsuit might not align with a hiking story, her sunglasses could still be fitting.
Therefore, \system enables object-level filtering by extracting all recognizable visual elements, forming a cutouts library prior to selection.

\subsubsection{Stage II: Element Selection and Clustering}
In this stage, we select (\textbf{DG1}) and cluster (\textbf{DG3}) cutouts associated with the story, leveraging the reasoning capabilities of an LLM---GPT-4o.
The input includes the user’s story description and a set of semantic labels representing visual elements.
To further satisfy \textbf{DG2}, elements classified under the ``Character'' category undergo part segmentation and pose estimation to generate manipulable cutouts of body parts. Stage II consists of the following steps:

\p{Label Selection.}
To address \textbf{DG1}, we prompt GPT-4o to select labels from the available semantic tags that maximize diversity while remaining relevant to the story content.
This process involves selecting assets to serve different roles in the narrative.
As shown in \autoref{fig:pipeline}, the selected elements fall into two categories:
(1) Central labels, which refer to objects explicitly mentioned in the story, and at least one of which is chosen for each object;
(2) related labels, which are relevant to the story but not directly stated.
The output is a JSON file containing the selected labels.
Any selected labels not found in the available label set are excluded.

\p{Role Classification.}
The second step classifies the selected labels into three predefined categories—characters, backgrounds, and accessories—which correspond to the fundamental roles that make up each scene (\textbf{DG3}).
This choice is informed by our formative study, which revealed that two primary clusters—characters and backgrounds—typically form the narrative backbone ($E_{1-2}$, $E_{4-5}$).
Characters, being central to the story, require the most manipulation time: “{\it I always spend most of my time on the characters.}” ($E_{4}$).
Backgrounds generally consist of larger, simpler elements that establish the setting, while accessories are smaller assets that enhance either the characters or the environment.
We prompt GPT-4o to separate the three categories based on the following rules:
Character layers are the attention center of the visual story, typically representing living beings and often derived from central labels.
Background layers consist of expansive scenery that fills the entire frame.
Accessory layers are independent objects that enrich the scene.
The output is a JSON file containing labels for each category.

\p{Layered Clustering and Character Parsing.}
To construct a semantic hierarchy, we prompt GPT-4o to cluster labels within each category into nested, layered groups (\textbf{DG3}).
Each cluster is assigned a name based on a summary of the labels of cutouts it contains (\textbf{DG3}).
For the first layer, labels are clustered based on their relationships to one another.
Each layer is then named by summarizing the category of the labels it contains.
Next, the upper layer is clustered based on the relationships between the names of the sub-clusters.
We prompt GPT-4o to repeat clustering until the relationships become vague, redundant, or when elements cannot be meaningfully grouped.
After clustering, we use Sapiens~\cite{khirodkar2024sapiens} to apply 2D part segmentation and pose estimation to character images.
Body part cutouts (\textbf{DG2}), along with their corresponding rotation centers, are generated to enable advanced editing of \system outputs.
At the end of Stage II, we obtain a nested JSON containing layered clusters of selected labels, each representing visual elements that align with the textual input.
For each character, body parts are represented as masks, and a JSON file containing body keypoints is generated.

We divided Stage II into three distinct tasks and regulated the intermediate results to ensure that the chosen visual elements exist among the visual elements provided by Stage I, and that the output is correctly formatted as a nested JSON.
This division is due to frequent hallucinations with single API calls that we identified in our early experiments. There, we provided examples and specified LLM roles and tasks; however, this resulted in intense selection of non-existent visual elements and incorrect categories.
The full prompts for Stage II are provided in \autoref{appendix:prompts}.

\subsubsection{Stage III: Asset Presentation}
The final stage concerns the visualization of selected assets for end-users to facilitate navigation and identification of desired elements (\textbf{DG3}).
The layered clusters are presented as a tree view with the names of clusters as nodes, alongside a canvas view rendering the visual assets.
The layout is arranged in a responsive grid, with elements ordered from left to right and top to bottom according to the sequence of clusters. The size of each visual element is varied based on a selection score.

\p{Asset Scoring.}
\system optimizes the size of a visual element to reflect a combination of three selection criteria: (1) $S_\text{div}$ for diversity in visual content, 
(2) $S_\text{cns}$ for consistency with the story, and (3) $S_\text{res}$ for image resolution.
In the following, we denote the visual element as $E_i$ and its associated cluster as $C$.

\paragraph{Diversity in content.}
\(S_\text{div}\) quantifies how a visual element differs from others within the cluster.
Suppose there are $|C|$ elements in cluster $C$. When $|C|>1$,\
\begin{equation}
S_\text{div}(E_i) = \frac{1}{|C|-1} \sum_{E_j\in C,\ i\ne j} \frac{1 - \text{cos}(e_i, e_j) + d(e_i, e_j)}{2}
\end{equation}
where $e_k$ is the normalized visual CLIP embedding for element $E_k$; $d(\cdot, \cdot)$ is the Euclidean distance; and $\text{cos}(\cdot, \cdot)$ denotes the cosine similarity, normalized to $[0, 1]$.
When $|C|=1$, i.e., the cluster contains $E_i$ only, then $S_\text{div}(E_i) = 0$.

\paragraph{Consistency with story.}
\(S_\text{cns}(E_i)\) measures the visual-text cosine similarity between $E_i$ and $C$, which is extracted from the original story description:
\begin{equation}
S_\text{cns} = \text{cos}(e_C, e_i),
\end{equation}
where $e_C$ is the textual CLIP embedding of the cluster $C$'s name, normalized to $[0, 1]$.

\paragraph{Resolution.}
The resolution score $S_\text{res}$ is designed to reflect the quality and clarity of a visual element based on its original resolution.
For each element, we assign a resolution score based on a min-max normalization of its resolution against the cluster:
\begin{equation}
S_\text{res}(E_i) = 
\begin{cases} 
\mathcal{N}(R_{E_i}, R_{\min}^{(C)}, R_{\max}^{(C)}) & \text{if } R_{\max}^{(C)} \neq R_{\min}^{(C)}, \\
1 & \text{if } R_{\max}^{(C)} = R_{\min}^{(C)},
\end{cases}
\end{equation}
where $R_{E_i}$ is the original resolution for $E_i$; $R_{\max}^{(C)}$ and $R_{\min}^{(C)}$ are the maximum and minimum resolutions of visual elements in the cluster $C$;
$\mathcal{N}(\cdot, \cdot, \cdot)$ is the min-max operator 
$\mathcal{N}(v, a, b)=(v-a)/(b-a) $.

\par\addvspace{0.5\baselineskip}%
Together, we combine the three selection criteria discussed above as the selection score:
\begin{equation}
S(E_i) = w_\text{div} \cdot S_\text{div}(E_i) + w_\text{cns} \cdot S_\text{cns}(E_i) + w_\text{res} \cdot S_\text{res}(E_i),
\end{equation}
where the weights \( w_\text{div} \), \( w_\text{cns} \), and \( w_\text{res} \) are adjusted according to the characteristics of the photo collection and user priorities.
In \system, these weights are set to .333 by default for equal importance.
We removed elements with the same score within the cluster to mitigate duplication, as SEM segmentation often yields visually similar segments, according to our early experiments.
In determining the height of an element,
\begin{equation}
h(E_i) = h_\text{0} + S(E_i) \cdot k,
\end{equation}
where $h_\text{0}$ is a constant base height and $k$ is an adjustable scaling factor.
The base height $h_\text{0}$ and scaling factor $k$ for characters are set higher than for other assets, as they are often the focus of a scene.

\subsubsection{\RR{Customization and Interoperability Support}}

\system is designed to be customizable and compatible to meet diverse needs in collage asset preparation.
Users may adopt a different category vocabularies and change the weighting rules for asset display size.
For specialized domains, they may heavily integrate visual elements that fall beyond generic labels, and they may lean on particular sides when weighting diversity, coherence, and resolution.
For instance, news production involves public figures and specific objects (e.g., \cite{politics, politics2}) that may fall beyond generic labels, while entertainment-oriented collage making prizes asset diversity over strict thematic adherence.
In addition, the exporter wraps up the curated assets with a preview collage, a JSON file declaring the semantic hierarchy and a folder of image cutouts for post-editing on other graphic tools.

\subsection{Interface}
\label{sec:interface}
A screenshot of the interface is shown in~\autoref{fig:workflow}.
The interface contains a source panel for user input, and a layer panel that presents curated cutouts and facilitates asset management with a linked tree view and canvas view.

\p{Source Panel: User Input.}
As the user imports a photo collection into the system, they can go through the source images to identify visual elements suitable for their narrative.
Then they can enter a story description into the input box and click the submit button to proceed.
The default text instructs the user to include visual elements as characters, background, and accessories for the story.
Upon clicking the submit button, the interface transitions to the layer panel.
visual assets are selected and grouped in the backend, then rendered on the layer panel.
The selection score is computed in the backend and influences the size of the layers on the canvas.

\p{Layer Panel: Asset Management.}
In the layer panel, the user can overview the collected visual assets, pick out, and lay out desired cutouts to create a collage scene.

The tree view displays the hierarchical semantic relationships of visual assets with a collapsible tree, where the leaf nodes represent individual cutouts.
Users can manage the visibility of these cutouts on the canvas through the side checkboxes.
They can also drag and drop nodes within the same cluster to adjust the layering hierarchy.
Additionally, clicking on a leaf node highlights the corresponding cutout on the canvas.

The canvas renders the initial presentation of selected visual assets.
It has a fixed width, and visual elements are rendered in order from left to right and top to bottom according to their clusters, potentially exceeding the canvas height.
Clicking on a visual element highlights the corresponding leaf node in the tree view.
Users can create basic static collages through direct manipulation of the cutouts, including copying, deleting, scaling, flipping, rotating, and boxes selecting of multiple elements.

%% file: sections/05_usageScenario.tex
\begin{figure*}[t]
    \centering
    \includegraphics[width=1\linewidth]{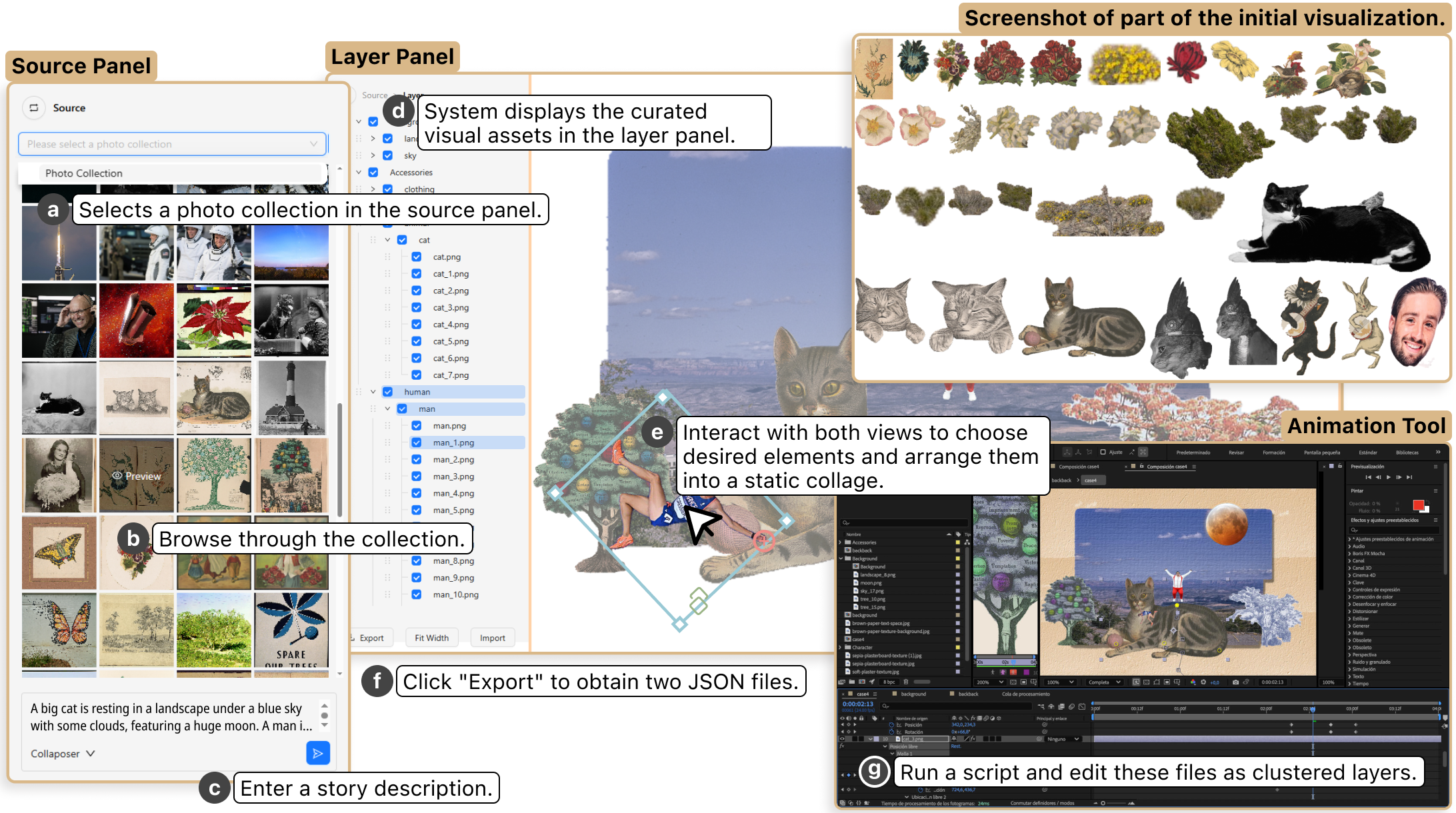}
    \caption{A user's workflow with \system. In the source panel, (a) the user selects a photo collection and (b) browses the collection. Then, the user (c) enters a story description in the input box and clicks the submit button. The system interface transitions to the layer panel when the preparation of visual assets is completed. (d) \system presents the visual assets in the layer panel. (e) User interacts with the tree view and the canvas view to pick up and arrange visual elements. After they are satisfied with the static collage story, (f) the user clicks the export button to obtain two JSON files. (g) They run a script and edit the collage story as clustered layers.}
    \Description{A user's workflow with \system. In the source panel, (a) the user selects a photo collection and (b) browses the collection. Then, the user (c) enters a story description in the input box and clicks the submit button. The system interface transitions to the layer panel when the preparation of visual assets is completed. (d) \system presents the visual assets in the layer panel. (e) User interacts with the tree view and the canvas view to pick up and arrange visual elements. After they are satisfied with the static collage story, (f) the user clicks the export button to obtain two JSON files. (g) They run a script and edit the collage story as clustered layers.}
    \label{fig:workflow}
\end{figure*}

\section{\RR{Usage Scenario}}
To illustrate how \system supports collage asset preparation, we walk through how \user, \RR{a representative 2D motion designer}, interacts with \system in one pass and produces a collage animation.
\autoref{fig:workflow} illustrates his entire workflow.

\p{Collect candidate cutouts with a story description.}
\user uploads a photo collection to the source panel (\autoref{fig:workflow}–a).
He has a general direction for the story and knows that the collection contains related visual elements, but he does not have a clear picture in mind.
\user is considering how different visuals can come together to convey his narrative while remaining open to creative possibilities.
He browses through the collection (\autoref{fig:workflow}--b) and writes a story in the input box (\autoref{fig:workflow}--c): ``\textit{A big cat rests under a blue sky with clouds and a huge moon; a man in white and red jumps on its back; another figure falls; trees and plants in the foreground.}''
The story indicates \user'  selection intent.
Here, the text includes fixed elements that are central to the narrative, such as {\it cat, man, woman, trees, plants}, {\it moon}, and {\it blue sky}, while abstract terms like {\it landscape} suggest additional elements that support the story.

After \user completes the input, \system leverages an LLM to extract labeled elements from photo collections and selects assets according to the input text.
The elements are then clustered and associated with a computed selection priority.
During this preparation, character assets are segmented into body parts with joint keypoints estimated for later animation.
Once preparation is completed, the collection preview switches to a layer panel with two linked views (\autoref{fig:workflow}--d): a hierarchical \emph{tree} on the left and a \emph{canvas} on the right.
In the canvas view, visual elements are laid out in a grid format.
Clustered elements are presented together, with their sizes adjusted based on selection criteria to enhance visibility and selection ease.
In the tree view, visual elements correspond to leaf nodes within a structured hierarchy of layers, providing an organized overview that supports efficient navigation.

\p{Immediately overview, pick, and lay out cutouts.}
\user obtains an overview of the visual assets and quickly identifies visual elements that catch his interest.  
He is satisfied with the assets, as they provide him with a rich selection to construct his story visually.
He interacts with both views to select desired elements and arrange them into a static collage (\autoref{fig:workflow}--e).
On the canvas view, he notices a piece of sky and a piece of ground that could serve as the background for the composition. 
He enlarges and positions them in the center of the canvas.  
Next, \user is drawn to a big cat and selects it, imagining it as the focal point of the collage.  
In the tree view, he notices a sub-cluster named ``cat'' under the character category, with a corresponding leaf node highlighted.
He opens ``characters/cat'', unchecks the cluster to hide all cats, and rechecks only the leaf corresponding to a large side-view of a cat so that just that cat appears on the canvas.
Next, he brings in a man in white and red and a falling athlete from ``characters/people''.
On the canvas, he spots interesting cutouts and uses direct manipulation to compose the scene: drag to position, use handles to scale/rotate, and order layer via the tree (dragging nodes up/down adjusts z-order).
Through the tree view, he navigates through concepts and hides unwanted clusters by unchecking.
With these operations, Lucas composes a balanced static collage within minutes.

\p{Post-edit.}
Lucas decides to animate the cat to increase the visual appeal.
Clicking the export button, he obtains two JSON files containing the asset metadata and the composed scene, along with a scene screenshot (\autoref{fig:workflow}--f).
To perform post-editing on the initial static collage, he runs an ExtendScript to import the files as clustered layers in Adobe After Effects (\autoref{fig:workflow}--g). Because character assets were pre-split with keypoints during preparation, they arrive as rigs with sensible rotation centers.
He stylizes the scene with paper texture and color tweaks, and then applies light stop-motion timing---the man in red and white jumps onto the cat, the cat’s back sags, and the athlete tumbles off—turning the static collage into a short, articulated animation.

\textit{Remark:} \user can prepare the cutouts effortlessly, thanks to the following advantages offered by \system.
\begin{itemize}
    \item In traditional workflows, \user would need to manually identify and extract desired visual elements from images. \system eliminates the distraction of unrelated objects by filtering out visual assets that do not align with the story.
    \item With traditional keyword-based search methods, \user would often need multiple attempts to account for elements overlooked in his initial query. By contrast, \system's ability to associate related elements supplements his incomplete textual descriptions, providing a rich set of cutouts to select optimal choices, enrich the scene, and get inspired by interesting combinations or elements. This is particularly beneficial when \user doesn’t have a clear vision of the visual story and is open to exploring possibilities.
    \item The quick overview of visual assets makes it easy for \user to detect and navigate through available options. The varied sizes of the visual elements help guide \user's attention to those with higher potential for selection.
    \item The exported files are structured and ready for advanced editing, with assets organized into layers, characters separated into parts, and key body points designated as rotation centers. This allows \user to adjust the visual effects of the clusters and create deformation and articulated animation on the characters.
\end{itemize}

%% file: sections/06_evaluation.tex
\begin{table*}[t]
\centering
\caption{An overview of participants in the user evaluation, including years of experience in creating collage-based stories (number of years), frequency of engagement, medium of collage practice, and professional domain.}
\label{t:evaluation_participants}
\begin{tabular}{rrlll}
\toprule
\textbf{ID} & \textbf{Number of Years} & \textbf{Frequency} & \textbf{Collage Practice} & \textbf{Profession} \\ 
\midrule
$P_{1}$ & 1 & yearly & web design & movie\\
$P_{2}$ & 3 & monthly & animated storyboard & animation\\
$P_{3}$ & 7 & yearly & fine art & design\\
$P_{4}$ & 7 & daily & pocket diary & management\\
$P_{5}$ & 5 & weekly & pocket diary & business\\
$P_{6}$ & 5 & seasonally & pocket diary & journalism\\
$P_{7}$ & 8 & seasonally & portfolio & design\\
$P_{8}$ & 4 & yearly & promotional video & visual arts\\
$P_{9}$ & 1 & monthly & promotional video & chemistry\\
$P_{10}$ & 9 & daily & commercial video & animation\\
$P_{11}$ & 14 & daily & commercial video & animation\\
$P_{12}$ & 13 & daily & art creation & media arts \\
\bottomrule 
\end{tabular}
\end{table*}

\section{\RR{User Evaluation}}
To evaluate Collaposer's effectiveness, we conducted a user study (\(N{=}12\)) addressing the following questions:
\begin{enumerate}[leftmargin=2.8em, topsep=0pt]
\renewcommand{\labelenumi}{\textbf{RQ\theenumi.}}
\item Selection quality: How well do the system-curated elements fulfill and expand upon a creator's stated intent?

\item Presentation effectiveness: How effectively does the system's presentation of elements support overview, navigation, and the discovery of meaningful connections?

\item Impact on creative practice: How does the integrated approach influence the creative workflow?
\end{enumerate}

\subsection{\RR{Methods}}
\RR{The study followed a within-subjects design in which each participant used \system and two baseline systems (see~\autoref{tab:baseline}).
This setup allowed us to isolate the effects of individual design components while keeping other factors constant.}
\RR{Quantitative data were collected from task duration, prompt attempts, and post-study questionnaires, while qualitative insights were drawn from screen recordings and semi-structured interviews.
The collected data were analyzed using non-parametric statistical tests for quantitative measures and thematic analysis~\cite{kiger2020thematic} for qualitative feedback.}

\subsection{Participants} 
We recruited 12 participants through social media and art communities, representing diverse collage-making experience and professional backgrounds (\autoref{t:evaluation_participants}).
Nine participants were collage amateurs ($P_{1\text{--}9}$) and three were professionals ($P_{10\text{--}12}$), defined as having received financial compensation for their collage work~\cite{rao2024scriptviz}.
Following~\cite{gunturu2025mapstory}, amateurs were compensated at US\$12 per hour and professionals at US\$50 per hour.

\begin{table*}[t]
\centering
\setlength{\tabcolsep}{4pt}
\caption{Comparison of \system and baseline systems.}
\label{tab:baseline}
\begin{tabularx}{\linewidth}{l X X l}
\toprule
\textbf{System} & \textbf{Asset Selection} & \textbf{Asset Presentation} & \textbf{Purpose}\\
\midrule
\system
& Infers central and related elements from story description
& Resizes and clusters assets based on the semantic hierarchy
& Evaluate the full pipeline capability (RQ3)\\
\midrule
Ablated-Select
& Extracts elements matching keywords in story description
& Same as \system
& Isolate selection effect (RQ1)\\
\midrule
Ablated-Present
& Same as \system
& Uses default height to resize assets
& Isolate presentation effect (RQ2)\\
\bottomrule
\end{tabularx}
\end{table*}

\subsection{Baselines} 

\autoref{tab:baseline} compares \system and two baseline systems in which its selection and presentation features are respectively ablated.
They share the same user interface, allowing users to switch between the systems with a dropdown menu (\autoref{fig:workflow}).
The \baselineA baseline generates candidate assets by matching the keywords mentioned in the user-input prompt without associating related elements (see~\autoref{appendix:prompts} for full prompts).
The \baselineB baseline presents assets without semantic clustering or adaptive resizing. 
Assets in \baselineB are displayed in a grid layout with the same height and random ordering, whereas assets in \system are arranged according to semantic clustering with varied sizes.

\subsection{Procedure}  
In the study, participants were tasked to create three stories from a given photo collection using \system and the two baselines respectively.
After completing the tasks, they finished a questionnaire and participated in an interview.
A session lasted about 90 minutes (Mean=91.33, SD=19.32).\footnote{The study protocol was approved by the institutional review board of the authors’ institution.} Each user session comprised three parts:

\textit{System Briefing (10 minutes).} Before the creation session, participants were given a demo and instructed to warm up with three systems to get familiar with the workflow and the interaction.

\textit{Open-ended Collage Story Creation (50 minutes).}
For each system, participants had up to 15 minutes to create a static collage story from scratch.
They assumed the role of collage enthusiasts, approaching collage creation with preliminary story ideas and an image library.
While \system can support diverse photo collections, including personal archives, client-provided references, and accessible image libraries, we employed a fixed photo set to avoid content variability influencing comparisons.
The order of the three systems was shifted to address ordering effects.
Immediately after each creation, the facilitator input the final prompt into other two systems and took a screenshot of the presentation.
This allowed for later comparison of curated assets from three systems using the same prompt, which facilitated participants’ evaluation in the post-study questionnaire.
Within 15-minute, participants tended to proceed with arranging elements on the canvas when the first preparation pass had yielded usable options.
\RR{This behavior suggests a satisficing strategy under time constraints: participants prioritized maintaining creative flow and completing a coherent scene within the allotted time, rather than optimizing through re-prompting.}
Consequently, the distribution of prompt attempts was highly skewed toward a single submission across the three systems.

\textit{Post-study Questionnaire and Interview (30 minutes).}
After completing the three stories, participants were asked to complete a questionnaire and participate in a semi-structured interview.

\subsection{Apparatus and Materials}
The study was run on a personal laptop with a 32 GB of memory for in-person attendees.
For remote attendees, the study was conducted on Zoom, using Chrome Remote Desktop to let participants control the system on the facilitator's computer.
A photo collection with 116 images was preprocessed and segmented into 2,427 visual elements on an Nvidia GeForce RTX~3060, requiring approximately 4~hours.
Control points and body parts for character layers were computed on an Nvidia GeForce RTX~4090, taking approximately 2--5~seconds per instance.
It took 5--20~seconds to select and present these visual elements for each user prompt.
Photos used in the user study were randomly selected from resources mentioned by the experienced creators in the formative study: the NASA Archives~\cite{nasa}, the Olympics Archive~\cite{olympic}, and the Library of Congress~\cite{congress}.

\subsection{Data Collection and Analysis}
We recorded the participants' prompt attempts to produce a satisfactory collage.
We define a \emph{prompt attempt} as a distinct submission of the story prompt within a system.
\emph{One-pass success} denotes a single submission followed only by canvas operations (selection, toggling, arranging) without re-prompting.
From screen recordings and interview notes, we coded participants’ reasons for re-prompting into three categories: (i) insufficient variety or missing key elements, (ii) overload from too many heterogeneous or irrelevant elements, and (iii) wording adjustments to better target desired elements.
As detailed in \autoref{appendix:questionnaire}, the post-study questionnaire asked participants to rate the systems on selection consistency, selection diversity, presentation, and system usability, based on a 7-point Likert scale.
We performed a Friedman test to compare the prompt attempts and rating distribution among the three systems, and conducted post-hoc Wilcoxon signed-rank test for system pairs to identify differences between the systems, using a Bonferroni-corrected alpha level of .0167 (.05/3).

The semi-structured interview aimed to better understand how the system assists participants in creating static collage stories, to explore the differences participants experienced using the three systems, and to anticipate potential applications.
The interviewer took notes and recorded audio during the interview, and verified the accuracy of the notes against the recordings.
We coded and analyzed the interview data via thematic analysis~\cite{kiger2020thematic}.
We categorized the data based on keywords within the scope of the three RQs and grouped similar ideas to identify recurring themes, which were then iteratively merged and refined.

%% file: sections/07_Results.tex
\section{Results}
Throughout the user evaluation sessions, the 12 participants created 36 static collage stories and tested 45 prompts with three systems.
Overall, \system outperformed the baselines in asset-story consistency, selection diversity, and usability.
It required fewer attempts on average, provided better visual alignment with user prompts, and was favored for its ability to inspire creativity.

\subsection{Iterations.}
All participants achieved one-pass success when using \system.
Under the two ablated baselines, some participants chose to iterate, because the initial element set was (i) too sparse or missing key elements, or (ii) too noisy/heterogeneous, with many distractors.
For example, some (\baselineA $P_{1}$, $P_{2}$, $P_{6}$, $P_{10}$; \baselineB $P_{10}$) altered the story or its components, while others refined terms to be more specific (\baselineA $P_{9}$; \baselineB $P_{1}$).
We analyzed the \emph{prompt attempts} per condition.
The Friedman test detected an overall effect of condition, $\chi^2(2)=8.38,\,p=.015$, but post-hoc Nemenyi tests did not reach significance for Collaposer vs.\ \baselineA ($p=.232$) or vs.\ \baselineB ($p=.866$; mean ranks: Collaposer $1.71$, \baselineA $2.38$, \baselineB $1.92$).
This does not contradict the one-pass result: the 15-minute per-story time limit induced a \textit{ceiling effect} and a \textit{zero-inflated} distribution (most trials across conditions had exactly one submission), limiting the sensitivity of pairwise tests despite descriptive differences.
Accordingly, we treat iteration outcomes as supportive—but not conclusive—evidence for RQ1–2 and triangulate them with questionnaire ratings and qualitative accounts below.

\begin{figure*}[t]
    \centering
    \includegraphics[width=\linewidth]{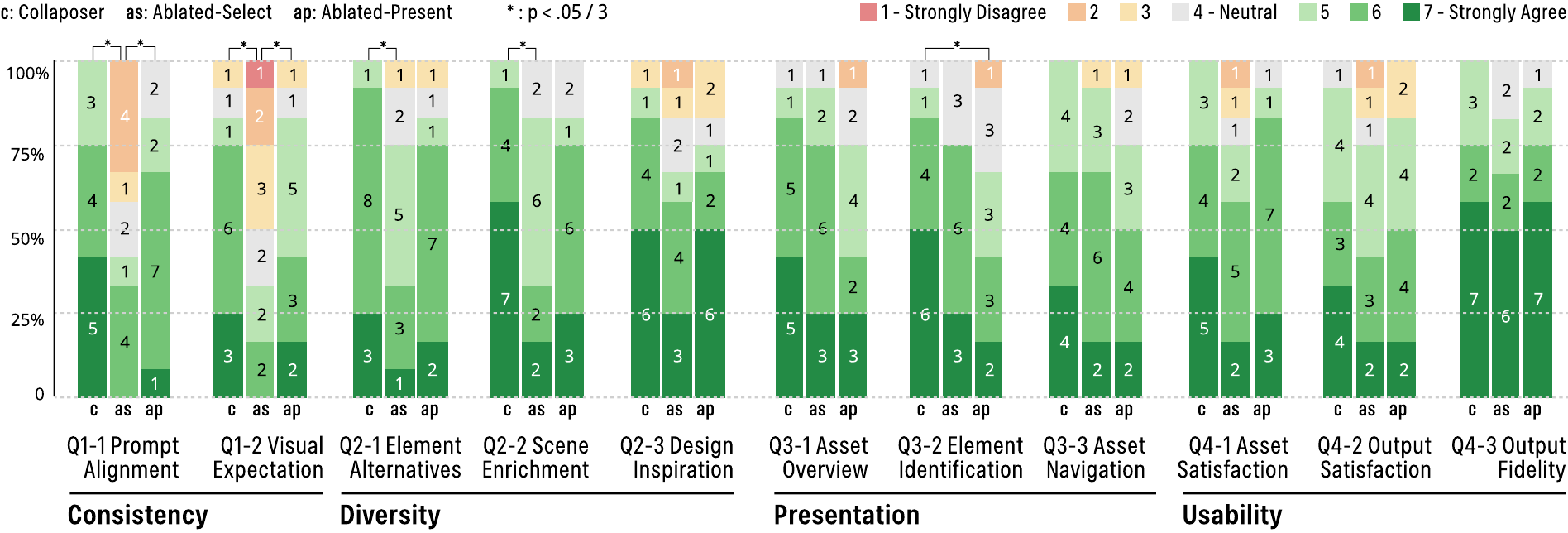}
    \caption{ \RR{User ratings across four evaluation dimensions—Consistency, Diversity, Presentation, and Usability—covering eleven question items (Q1–Q4) for three system variants: \system (c), \baselineA (as), and \baselineB (ap). Ratings were collected on a 7-point Likert scale (1 = Strongly Disagree, 7 = Strongly Agree). Asterisks (*) indicate statistically significant differences in mean ratings (p < .05 / 3).
    Overall, the results indicate that \system supports more effective story-aligned asset selection and presentation compared to the baselines. \baselineA shows the weakest asset-story consistency, often failing to provide relevant elements and occasionally including unrelated ones. \baselineB delivers the least satisfactory presentation results, though it has a relatively smaller impact on system usability.}}
    \Description{User ratings across four evaluation dimensions—Consistency, Diversity, Presentation, and Usability—covering eleven question items (Q1–Q4) for three system variants: \system (c), \baselineA (as), and \baselineB (ap). Ratings were collected on a 7-point Likert scale (1 = Strongly Disagree, 7 = Strongly Agree). Asterisks (*) indicate statistically significant differences in mean ratings (p < .05 / 3). Overall, \system outperforms baselines across all dimensions, demonstrating its ability to select diverse visual elements that align with the story and effectively present them for collage creation. In contrast, \baselineA shows the weakest asset-story consistency, often failing to provide relevant elements and occasionally including unrelated ones. \baselineB delivers the least satisfactory presentation results, though it has a relatively smaller impact on system usability.}
    \label{fig:rating}
\end{figure*}

\subsection{Questionnaire.}
The analysis and data visualized in \autoref{fig:rating} demonstrate that \system consistently outperformed the baselines in aligning visual assets with users’ selection intent.
All participants confirmed that \system’s assets matched their prompts (Q1-1), and 11 out of 12 found them consistent with the story content (Q1-2).
Post-hoc \textsc{Wilcoxon Signed-Rank Tests} with \textsc{Bonferroni}-corrected alpha .0167 showed significant differences between \system and \baselineA in Q1-1 and Q1-2.
By contrast, \baselineA frequently provided irrelevant elements ($P_{2-4}$, $P_{10}$, $P_{12}$) or included unrelated assets ($P_{6}$, $P_{7}$, $P_{10}$), reducing both selection consistency and diversity. \system also offered richer alternatives to complement participants’ selections (Q2-1, Q2-2) and inspired visual expression (Q2-3), with a significant advantage over \baselineA in Q2-1.
\baselineB performed moderately in diversity but showed weaknesses in presentation.

Participants could quickly overview, locate, and navigate assets with \system's presentation (Q3-1 to Q3-3), and post-hoc tests indicated significant differences between \system and \baselineB in Q3-2.
\baselineB showed the poorest presentation performance, though its impact on usability was smaller.

For overall usability, all participants agreed that \system supported preparing visual assets for collage-based storytelling (Q4-1) and could serve as intermediate outputs (Q4-3).
Eleven out of 12 participants created satisfactory static collages.
The remaining participant, $P_{10}$, expressed a neutral stance, noting that while \system's image segmentation precision did not meet commercial standards, it was adequate for personal creations on social media.

\begin{figure*}[t]
    \centering
    \includegraphics[width=1.0\linewidth]{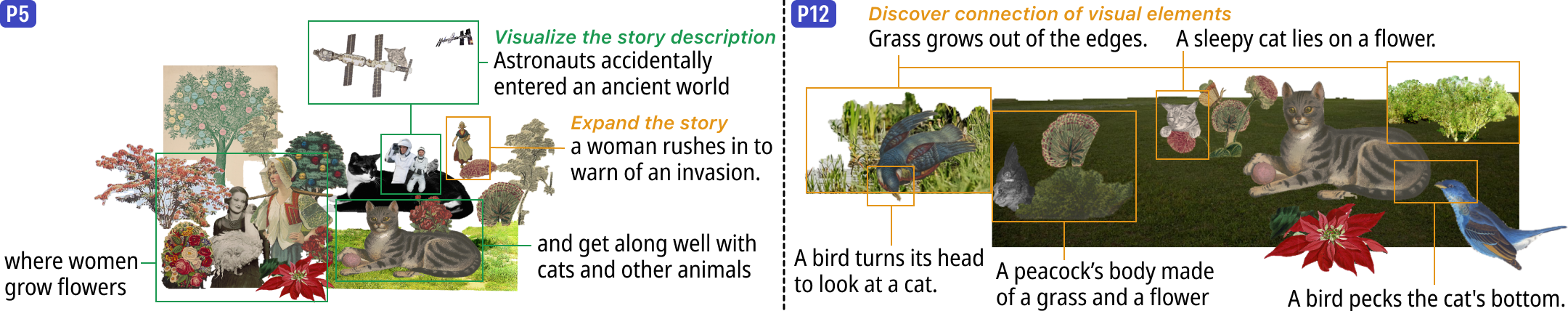}
    \caption{\system's selection aligns with P5's mental visualization of the story and enriches some details. P12's story created with \system starts from a broad description ``different animals are playing on the playground'' and adds details through the composition of visual elements.}
    \Description{P12's story created with \system starts from a broad description ``different animals are playing on the playground'' and adds details through the composition of visual elements.}
    \label{fig:p5-12}
\end{figure*}

\subsection{Semi-structured Interview}
\label{interview}
 We then conducted interviews to examine participants’ rating rationales and \system’s impact on their creative practices.

\p{Selection Consistency and Story Enrichment (RQ1).}
Participants found visual assets generated by \system to be consistent with their story descriptions and to offer diverse options.
Compared to \baselineA, \system was able to provide more potentially related elements to enrich the scene ($P_{1-4}$, $P_{6-7}$, $P_{9-12}$) and was less likely to miss necessary story components ($P_{2-4}$, $P_{6-7}$, $P_{9-12}$).
\system was able to ``{\it supplement elements not explicitly mentioned in prompts}'' ($P_{9}$), ``{\it present segmented parts of the same image}'' ($P_{7}$), ``{\it offer elements spanning foreground, background, and characters}'' ($P_{5}$), and ``{\it generate rich background cutouts}'' ($P_{4}$).
Within a manageable scope for creators, having more elements is preferable since supplementing missing assets is costly.
The selection capability inspired users to add more details to the collage story ($P_{1-12}$) and expand the narrative ($P_{6}$, $P_{8}$, $P_{11-12}$),
rather than altering the story description (\baselineA $P_{1}$, $P_{2}$, $P_{6}$, $P_{9-10}$, \baselineB $P_{1}$, $P_{10}$) or adjusting the story direction based on the curated assets (\baselineA $P_{2}$, $P_{8-9}$, $P_{12}$).
``{\it \system provided me with an additional house, which is related to the woman, adding her background information and depth to the photo. In contrast, when I requested a biker, \baselineA gave me a biker without a bicycle, so I let the character pat a moon as a ball instead,}'' $P_{11}$ explained.
Overall, \system was judged to effectively curate visual assets that support story content and provide a rich space for further selection.

\p{Presentation with Clustering and Size Variability (RQ2).}
Presentation of clustered visual assets was reported to simplify the comparison of similar objects for informed selection ($P_{1}$, $P_{3}$, $P_{5}$, $P_{8}$), to help envision interactions between visual elements ($P_{12}$), and to create an organized impression ($P_{6}$, $P_{10-11}$).
``{\it I first remove some oddly shaped fragments and those with low clarity, then collage the background. After that, I select individual objects that interest me and combine them to create a narrative between them (see \autoref{fig:p5-12}).
Grouping similar elements and the order of arrangement align closely with my creation process,}'' $P_{12}$ noted.
Although \system and \baselineB adopt the same selection technique, participants perceived the assets provided by \baselineB as less related ($P_{4}$, $P_{6-7}$) and more competitive ($P_{1}$, $P_{6-7}$).
As $P_{1}$ pointed out ``{\it Without categorization, it feels overwhelming.}''
While half of the users did not notice differences in element sizes, others recognized this variation demonstrated potential to help with selection and influence the narrative direction ($P_{1}$, $P_{3-5}$, $P_{8-9}$, $P_{12}$).
``{\it Altered sizes create a more flexible and creative mood}'': ($P_{3}$);
``{\it The big cat and small human inspired me to create a world with big cats and small people. It's interesting and unexpected}'' ($P_{5}$);
``{\it With various sizes, it's easier to select what I want. The visualization in \baselineB is too dense, and I felt overwhelmed.}'' ($P_{12}$).
Bigger elements also grab creators' attention:
``{\it I noticed some bigger assets align with my target selection, especially the characters}'' ($P_{9}$).
\p{Rapidly transforming from Conceptual to Visual Story (RQ3).}
By automating the preparation of visual assets, \system allowed creators to start from a concept and quickly review and arrange a visual story based on a set of curated cutouts ($P_{1-12}$), eliminating the time-consuming steps of searching for and segmenting images ($P_{2}$, $P_{6}$, $P_{7}$), and enabling efficient selection and extraction of relevant elements ($P_{5}$, $P_{8}$).
It was especially useful when the story was open-ended, supporting ideation ($P_{4-5}$,$P_{9-12}$) and development of the narrative ($P_{1-12}$).
For example, $P_{1}$ started from a rough story concept, ``{\it a girl walks through different places}'', and was able to quickly assess and organize a visual story using the cutouts.
While some participants mentally visualized the scene before receiving the elements (see~\autoref{fig:p5-12}), others ($P_{1}$, $P_{4}$, $P_{7}$, $P_{12}$) relied on serendipity, gathering rich material and laying out visual stories based on elements that interested them.
Users found inspiration to develop stories from visual elements ($P_{1-12}$) and their combinations ($P_{1-5}$, $P_{11}$, $P_{12}$~\autoref{fig:p5-12}).
Compared to manual scanning, which easily leads to overlooking certain materials and requires mental effort to remember and connect stories, \system assisted creators to uncover overlooked details ($P_{3}$, $P_{4}$).

\p{Resolving Ambiguity in Story Conceptualization with Rich Collages (RQ3).}
Prior HCI research has emphasized that design materials actively shape ideation and outcomes, as materials ``talk back'' to designers and guide decision-making through their properties~\cite{Pradhan2025OlderAdults, Giaccardi2015MaterialsExperience, Tholander2012Agency, Schon1992}.
In our context, visual assets provided by \system function as design materials for storytelling and shape the space of story creation.
During creation, the visual assets helped participants concretize potential scenes and informed decision-making, including both affirmative and negative judgments.
The absence of sufficient assets signaled that a certain concept could not be sustained, whereas the presence of particular elements served as cues for narrative development.
Even when a story ultimately incorporated only a few cutouts, the exposure to a broader pool of assets sparked inspiration that might not otherwise arise.
In some user study cases, the collection lacked specific intended elements.
For example, $P_2$ needed paper planes, but no metallic planes existed in the system, necessitating a change in the story. 
This feedback led to adjustments to the narrative or the expansion of the asset collection.
$P_6$ chose to exclude dogs due to dissatisfaction with the available options.
With \system, creators could quickly verify whether a proposed idea was feasible with the existing materials, allowing them to plan and refine stories more effectively.
As $P_4$ reflected, the system allowed for rapid assessment of how many relevant assets are available to serve as visual clues in the story.

\begin{figure*}[t]
    \centering
    \includegraphics[width=1\linewidth]{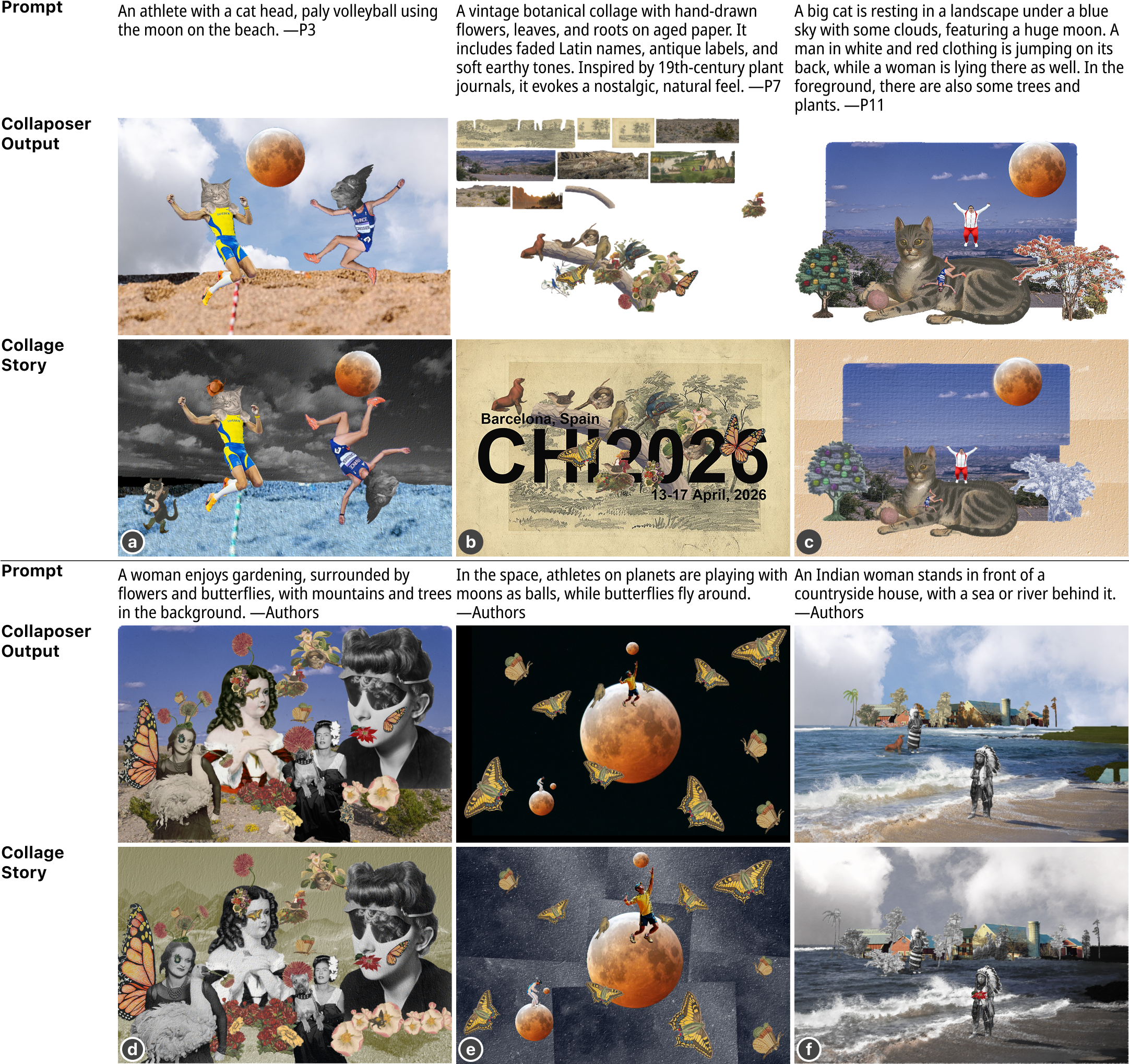}
    \caption{Collage stories created with outputs of \system.}
    \Description{Collage animations created with outputs of \system.}
    \label{fig:gallery-animated}
\end{figure*}
\p{Potential Applications (RQ3).}
Participants anticipated the potential application of our system's output.
With personalized photo collections, social media posts were a popular choice, including ``{\it artistic annual review vlogs}'' ($P_{4}$), ``{\it short videos on TikTok that feature layers appearing in sync with the music rhythm}'' ($P_{2}$), ``{\it dream maps on Instagram}'' ($P_{10}$), modern art expression($P_{12}$), and personal posts created just for fun ($P_{5}$, $P_{8}$).
In addition to entertainment, by adding text, layer effects (such as textures), precise cutouts, and even simple animations, collages could be transformed into posters that effectively convey information ($P_{5-6}$, $P_{8-9}$).
The richness of visual assets and the speed of creation were thought to make \system ideal for animation storyboards ($P_{6}$), concept maps ($P_{6}$, $P_{8}$), problem statements in portfolios ($P_{7}$), and mood boards ($P_{8}$).
As collages are considered creative and beloved by the general public, some participants said that they have the potential to be used for innovative advertising ($P_{7}$, $P_{9}$, $P_{11}$) and education ($P_{9}$, $P_{11}$).


\p{\RR{Expanding \system’s Output to Collage Animations (RQ3).}}
To showcase the effectiveness of \system and assess how well its output integrates with commercial graphic design and animation tools, we created six collage animations (\autoref{fig:gallery-animated}) with outputs from \system made from the same photo collections for user evaluation.
$S_{a-c}$ were initially crafted by $P_{3}$, $P_{7}$ and $P_{11}$ during the user evaluation.
\system provided background assets even if they were not explicitly mentioned in the prompts ($S_{a}$, $S_{d}$, $S_{f}$), along with a variety of accessories related to the story to enhance the scene ($S_{b-d}$, $S_{f}$).
In Story f, the description mentions only key elements: ``an Indian woman'', ``a countryside house'', and ``a sea or river''.
\system enhanced this by adding skies, sea animals, and trees, enabling the creator to build a more complete scene.
We created the static scenes using \system and then made advanced edits in Adobe After Effects, as detailed in \autoref{appendix:advancededit}.

%% file: sections/08_discussion.tex
\section{\RR{Discussion}}
\RR{\system introduces a story-driven approach for asset preparation in digital collage creation.
By automating related elements extraction and organizing them based on semantic labels, it demonstrates how LLM-based assets retrieval can supplement user expression, support serendipitous exploration, and facilitate nonlinear iteration.
This section distills the design implications derived from \system, highlighting how they may inform future generative AI–powered creativity-support tools.}

\p{\RR{Supplementing Underspecified Components with Semantic Inference:}}
\RR{Rather than assuming precise input queries, which in many cases place heavy cognitive demands on users~\cite{Subramonyam2024BridgingGulf}, \system enables ``divergent selection'', translating a vague story description into a space of visual concepts semantically relevant to the story.
Leveraging LLMs, \system retrieves not only central elements explicitly mentioned in the story descriptions, but also related elements inferred to be plausible scene components (i.e., characters, backgrounds, and accessories) for the target story. 
This approach effectively retrieves relevant elements to enrich the story (\autoref{interview}--RQ1).
With \system, users navigate and select visual assets from a broad pool of candidates without needing to articulate detailed textual specifications.
Their attention is therefore shifted from textual articulation to visual exploration, reconstructing the early stage of collage creation into a process where ideation relies less on linguistic precision and more on navigating a diverse, LLM-expanded visual space.
More broadly, this paradigm suggests how content-creation tools may augment underspecified user intentions with model-learned priors, such as common components associated with a given creation type, to meaningfully expand the design space for discovery and inspiration.}

\p{\RR{Supporting Serendipitous Discovery in LLM-powered Retrieval:}}
\RR{LLM-powered search is inherently unpredictable~\cite{yampolskiy2024ai}, which may retrieve visual elements deemed relevant by LLM but unexpected for users.
However, such unpredictability introduces opportunities for serendipitous inspiration, especially in early-stage ideation.
In \system, we implement a human-AI relay where the AI pre-filters candidates, leaving the final judgment to users.
The returned candidates serve as provocations that broaden the user’s exploration space.
We find users reconsider, refine, and expand their initial stories after encountering unexpected visual elements.
For instance, $P_1$ wrote ``a cat falls in love with human'' and obtained several hand elements.
While this mismatched their anticipation of full human body only, $P_1$ found such detail introduced intimacy between cats and humans, and integrated two hand elements to the final collage.
The serendipitous findings help users explore novel story directions and produce out-of-the-box design.
Accordingly, future AI-powered creativity-support tools, especially those for compositing artifacts such as world building in games, may consider fostering serendipitous discovery by tolerating loosely relevant candidates.}
\p{\RR{Embodying Semantic Concepts into Manipulable Alternatives for Nonlinear Iteration:}}
\RR{Nonlinear creative iteration rarely unfolds through a single, fully formed intention; instead, creators’ goals shift, branch, and evolve throughout the process~\cite{creativeDesign}.
Meanwhile, nuanced user intentions are often difficult to fully verbalize~\cite{Lazar2018MakingExpression, Kreminski2024Intent}.
To support nonlinear exploration guided by underarticulated user intentions, \system embodies each semantic concept into multiple manipulable alternatives, translating abstract descriptions into candidate visual elements.
In this way, users can explore possibilities through direct manipulation---selecting, replacing, or combining alternatives on the canvas.
This approach may enable a nonlinear iteration by shifting cognitive effort from verbalizing nuanced creation intentions to evaluating potential visual element options, allowing users to refine collage compositions at the object level.
As a result, decision-making becomes more grounded, exploratory, and responsive to emergent ideas.
Future tools could also consider representing a scene not just as a single output, but as a set of manipulable alternatives for each semantic concept.
The sources for these alternatives may include creation histories, AI-generated content, user-provided collections, or resources from the internet.
Without requiring users to repeatedly modify textual descriptions, such decomposition externalizes the evolving and often ambiguous intentions to better supports nonlinear iteration.}

%% file: sections/09_limitationsFutureWork.tex
\section{\RR{Limitations and Future Work}}
\RR{We next discuss limitations and directions for future research.}
\p{\RR{Text-based Intention Expression.}}
Although \system made it easier for creators to obtain assets that support story development through text, we still observed instances of misalignment between prompts and users’ intended selections.
This was reflected in the questionnaire results, where scores for Q1-2 were slightly lower than those for Q1-1 (\autoref{fig:rating}).
Future research could develop richer mechanisms to elicit or support creators’ selection intent, addressing the limitations of relying solely on text-based expression.

\p{\RR{Transparency and User Trust.}}
\RR{While \system’s semantic inference broadens creators’ exploration space, we find users confused about why certain assets are suggested and how the system interprets their story descriptions.
Prior work indicates that even when AI-powered tools reduce cognitive load and enhance user experience, users often express concerns about the explainability of model operations and data usage\cite{xu2024trust}.
Similar concerns arise when \system influences ideation through behind-the-scenes inference.
Future work could explore ways to make semantic inference more transparent---such as providing explanations or adjustable inference scope---helping creators maintain a clear sense of agency while benefiting from the expanded creation space.}

\p{From Single-Pass to Iterative Preparation.}
\RR{\system currently performs a single-pass asset preparation based on story descriptions, but collage storytelling often involves continuous refinement between asset selection and composition.
To better scaffold iterative refinement, future work could support dynamic organization of asset clusters, allowing users to regroup or rename clusters and reconfigure their hierarchy by facets such as color, artistic era, or style as their creation focus shifts.
Another direction is to enable local precision loops, such as interactive masking tools for re-segmentation of images.
These targeted functions would allow users to fix regional adjustments without restarting the entire pipeline.}

\p{User Evaluation.}
Our user study involved twelve participants with varying creative backgrounds, allowing us to capture a range of perspectives on how \system\ supports story-driven collage making. 
However, the short study duration limited our understanding of long-term engagement.  
Future work could include longitudinal studies to observe how users integrate \system\ into their creative workflows over time and across different project stages.  

%% file: sections/10_conclusion.tex
\section{Conclusion}
In this paper, we introduce \system, a tool that transforms photo collections into organized visual assets to be used for telling stories with collages, both statically and dynamically.
Informed by practitioners' workflows identified through a formative study, \system leverages an LLM-based pipeline to select, separate, cluster, and present visual elements embedded in images that align with a given story description.
These visual assets are drawn either directly from the prompt or inferred from abstract related concepts.
A user study demonstrated that \system effectively prepares diverse visual assets relevant to story content, while enabling users to quickly select elements for collage-based storytelling.
We further demonstrated the utility of \system by creating collage animations with its output of static collages.

%% file: sections/11_appendix.tex
\appendix
\vspace{20pt}
\section{Visual Assets Selection and Clustering}
\label{appendix:prompts}
The system prompt of \system guides an LLM model to infer central and related elements from a story description.
\begin{lstlisting}
### Role
- You are a selector of visual assets. You need to select visual assets based on the story.

### Instructions
- The visual assets should be relevant to the user input and should be diverse to offer options and enrich the scene. Choose assets as characters, backgrounds, and accessories.
- For assets mentioned in the story, use the direct labels; make sure to find the most appropriate label for each. For example, if you can't find "boy", consider using "man" or similar labels.
- For objects that are not explicitly mentioned but are relevant, select appropriate related labels. For example, if the story is about a garden, you can select related objects that are likely to appear in a garden, such as tree, flower and butterfly as accessories, sky and grass as background.
- The visual assets available are: [labels_list]

### Example
- user_input: A sunny day, a boy and a dog are playing in the park.
- labels_list: [boy, girl, woman, man, dog, cat, park, sky, sun, grass, flower, tree, frisbee, car, skirt, bank, house, building, street, road, sidewalk, bench, swing, slide, ball, kite, umbrella, hat, sunglasses, cloud, moon, stars, mountain, river, lake, ocean, beach, sand, boat, ship, airplane, helicopter, train, bus]
- ai_output:
    - Direct labels: [boy, dog, park]
    - Related labels: [sky, sun, cloud, grass, tree, flower, frisbee, ball, sunglasses]
\end{lstlisting}

\baselineA's prompt guides the model to extract elements matching keywords in a story description.
\begin{lstlisting}
### Role
- You are an assistant for visual assets preparation.

### Instructions
- The visual assets should be mentioned in the story.
- The visual assets available are: [labels_list]

### Example
- user_input: A sunny day, a boy and a dog are playing in the park.
- labels_list: [boy, girl, woman, man, dog, cat, park, sky, sun, grass, flower, tree, frisbee, car, skirt, bank, house, building, street, road, sidewalk, bench, swing, slide, ball, kite, umbrella, hat, sunglasses, cloud, moon, stars, mountain, river, lake, ocean, beach, sand, boat, ship, airplane, helicopter, train, bus]
- ai_output: [boy, dog, park]
\end{lstlisting}

This prompt guides the model to construct a semantic hierarchy of the selected labels.
\begin{lstlisting}
### Step 1: Asset Classification
- Role: You are a classifier of visual assets.
- Instruction: Classify the visual assets into three categories: Characters, Accessories, and Background. The background is expected to be huge scenery that fills the entire frame.
- Example:
  - labels_list: 
      - Direct labels: boy, dog, park
      - Related labels: sky, sun, cloud, grass, tree, flower, frisbee, ball, sunglasses
  - AI output:
      - Character: [boy, dog]
      - Accessories: [frisbee, ball, sunglasses, sun, cloud, grass, tree, flower]
      - Background: [park, sky]

### Step 2: Asset Clustering
- Role: You are a cluster of visual assets.
- Instruction: Group visual assets within each category into subcategories based on their relationship to each other. Keep clustering until it no longer makes sense.
- Example:
  - labels_list: 
      - Character: [boy, dog]
      - Accessories: [frisbee, ball, sunglasses, sun, cloud, grass, tree, flower]
      - Background: [park, sky]
  - AI output:
      - Character: [boy, dog]
      - Accessories:
          - dog toy: [frisbee, ball]
          - human belongings: [sunglasses]
          - environment: [sun, cloud]
          - plant: [grass, tree, flower]
      - Background: [sky, park]
\end{lstlisting}

\section{Post-Study Questionnaire}
\label{appendix:questionnaire}
The post-study questionnaire is based on a 7-point Likert scale 
(1 = strongly disagree, 4 = neutral, 7 = strongly agree).
    
    
    

\begin{enumerate}[label=Q \negthinspace\arabic*., left=0pt]
    \item \textbf{Visual Elements - Story Content Consistency}
    \begin{enumerate}[label*=\arabic*.]
        \item The selected visual assets align with my prompt.
        \item The selected visual assets meet my expectation of story content.
    \end{enumerate}
    
    \item \textbf{Visual Elements - Element Diversity}
    \begin{enumerate}[label*=\arabic*.]
        \item For fixed elements, the selected visual assets provide enough diversity to offer options.
        \item For fixed elements, the selected visual assets enrich the scene with variety.
        \item Some visual elements inspire me regarding the visual expression of the story.
    \end{enumerate}
    
    \item \textbf{Presentation}
    \begin{enumerate}[label*=\arabic*.]
        \item I can quickly obtain an overview of the visual elements I have with the presentation.
        \item I can easily locate the visual elements I want within the presentation.
        \item I can easily navigate through the visual elements with the presentation.
    \end{enumerate}
    
    \item \textbf{Usability}
    \begin{enumerate}[label*=\arabic*.]
        \item The system helped me prepare visual assets for collage-based storytelling.
        \item I can use the system to create satisfactory static collages.
        \item The system’s outputs can be intermediate results of collage-based storytelling.
    \end{enumerate}
\end{enumerate}

\section{How does \system output support dynamic control of animation and layout?}
\label{appendix:advancededit}
We create static scenes with \system and export them for use in After Effects, where each cluster becomes a composition and each visual element is placed as a layer according to its original position.
Invisible layers are also imported but remain hidden.
Character layers are replaced by body parts, with key body joints set as rotation centers. 
Each character becomes a separate composition.

\p{Layout adjustment and visual enhancement.}
First, we perform layout adjustments, including duplicating layers, modifying layer orders (e.g., the mountain in $S_d$, and the flower in $S_f$ are initially blocked), and fine-tuning layer positions within clusters. 
In $S_b$, an animated poster, we add texts between the butterfly cluster and other clusters, and move the birds and sea lion in front of the text.
We then apply visual adjustments by modifying the color of clusters and adding different paper textures to unify tones and enhance contrast between characters and their surroundings.

\p{Animation design and production.}
Clusters guide animation planning.
In $S_b$, butterflies serve as a transition.
They flap outward from the center, revealing the poster underneath.
In the final scene, two butterflies flap while a sea lion moves from the ``B'' to the ``e'' in ``Barcelona'' (stop-motion).
In $S_d$, we animate the plant and insect clusters.
Butterflies flutter, bees buzz, and flowers sway.
Other elements remain static.
In $S_f$, we visualize wind.
Ground and building clusters remain still.
Trees sway in sequence, clouds drift, a boy’s feathered headpiece flutters as he shakes his head, and a flower falls from his hands onto the beach.
For character animation, we inspect the segmented parts, refine the body parts, create smaller parts, and add rigging or puppet pins.

\p{Layout adjustment and visual enhancement.}
First, we perform layout adjustments, including duplicating layers, modifying layer orders (e.g., the cat in $S_a$, the mountain in $S_d$, and the flower in $S_f$ are initially blocked), and fine-tuning layer positions within clusters. 
In $S_b$, an animated poster, we add texts between the butterfly cluster and other clusters, and move the birds and sea lion in front of the text.
We then apply visual adjustments by modifying the color of clusters and adding different paper textures to unify tones and enhance contrast between characters and their surroundings.

\p{Animation design and production.}
Clusters guide animation planning.
In $S_a$, a man on the right kicks the moon like a ball, demonstrating leg rotation.
The moon deforms on impact and flies toward the man on the left.
Simultaneously, the man on the right rotates 360 degrees due to the force.
The moon hits the man on the left's face, both deform, and the moon returns to the right.
In $S_b$, butterflies serve as a transition.
They flap outward from the center, revealing the poster underneath.
In the final scene, two butterflies flap while a sea lion moves from the ``B'' to the ``e'' in ``Barcelona'' (stop-motion).
In $S_c$, each non-character layer enters the frame from outside using stop-motion.
After reaching his position, the red man jumps on the cat, causing its back to sag and the blue man to fall off.
In $S_d$, we animate the plant and insect clusters.
Butterflies flutter, bees buzz, and flowers sway.
Other elements remain static.
On planets in $S_e$, two athlete swings at the moon like a badminton shuttle or bounces it like a ball.
Each hit makes the moon orbit the planet.
Butterflies float motionlessly while mysterious stars flicker in the background.
In $S_f$, we visualize wind.
Ground and building clusters remain still.
Trees sway in sequence, clouds drift, a boy’s feathered headpiece flutters as he shakes his head, and a flower falls from his hands onto the beach.

For character animation, we inspect the segmented parts, refine the body parts, create smaller parts, and add rigging or puppet pins.
For instance, in $S_a$, to support the kicking motion, the left athlete’s arm has been segmented, with the shoulder assigned as the rotation center.
The right athlete’s arms, upper and lower legs, and feet have been fine-tuned, with each part segmented and assigned to the appropriate joints (shoulder, hip, knee, ankle).
In $S_e$, the athletes’ torsos and legs have been separated to support crouching through deformation.
We apply puppet pins at the joints.